\documentclass[conference]{IEEEtran}

\IEEEoverridecommandlockouts
\usepackage{cite}
\usepackage[dvipsnames]{xcolor}
\usepackage{hyperref}
\usepackage{subcaption}
\usepackage{graphicx}
\usepackage{braket}
\usepackage{ifthen}
\usepackage{booktabs}
\usepackage{wheelchart}
\usepackage{pgfplots}

\newboolean{isfull}
\setboolean{isfull}{true}
\newcommand{\appcite}[1]{%
    \ifthenelse{\boolean{isfull}}{App.~\ref{#1}}{the full paper}}

\begin{document}

\ifthenelse{\boolean{isfull}}%
    {\title{Bridging the Quantum Divide: A Learning-Centric Quantum Hackathon for Underrepresented Students (Extended Version) \thanks{This project was funded in full by the CISE-Atlantic Catalyze and Facilitate Sponsorship Program (2025).}}}%
    {\title{Bridging the Quantum Divide: A Learning-Centric Quantum Hackathon for Underrepresented Students \thanks{This project was funded in full by the CISE-Atlantic Catalyze and Facilitate Sponsorship Program (2025).}}}%

\author{\IEEEauthorblockN{Fahimeh Bayeh}
\IEEEauthorblockA{\textit{Dalhousie University}\\
Halifax, Canada \\
fahimeh.bayeh@dal.ca}
\and
\IEEEauthorblockN{Linh Dinh}
\IEEEauthorblockA{\textit{University of Ottawa}\\
Ottawa, Canada \\
ldinh052@uottawa.ca}
\and
\IEEEauthorblockN{Dongho Lee}
\IEEEauthorblockA{\textit{Unaffiliated}\\
Seoul, South Korea \\
fredldh@gmail.com}
\and
\IEEEauthorblockN{Scott Wesley}
\IEEEauthorblockA{\textit{Dalhousie University}\\
Halifax, Canada \\
scott.wesley@dal.ca}
}

\maketitle

\begin{abstract}
    This paper describes the design and implementation of a two-day quantum hackathon for underrepresented high school students in Nova Scotia, Canada.
    The first day of the hackathon is spent introducing students to quantum computing through hands-on activities, whereas the second day teaches students to apply this knowledge through guided challenges.
    Both days are informed by the theory of mastery learning and specification grading, with the full curriculum being crafted within the Integrated Course Design framework.
    This requires identifying situational factors unique to our target demographics, from which we develop learning outcomes, and then work backwards to a full curriculum with educative assessments.
    A novel aspect of our hackathon is that all circuit simulations are performed within Quirk: a decision based on best practices in computer science education.
    Based on feedback from students, we conclude that our hackathon successfully introduced students to the basics of quantum computing, and was able to reach most of our target demographics.
\end{abstract}

\begin{IEEEkeywords}
    % The IEEE has a keyword theasurus.
    Quantum computing,
    quantum simulation,
    computer science education,
    physics education,
    curriculum development,
    educational technology,
    equal opportunities,
    gender equity
\end{IEEEkeywords}

\section{Introduction}

In 2024, the United Nations declared 2025 to be the International Year of Quantum, and made a global call for the development of a diverse and inclusive quantum workforce~\cite{DNR2025}.
This echoes the policies of many governmental organizations which are now invested in the ``second quantum revolution'' and have identified quantum education as an important priority~\cite{ISED2022,QF2016,DISR2023,SQIS2022,R2025}.
It has been argued that for true quantum readiness, education must start as early as in secondary school~\cite{PFKRM2020}.
Indeed, many countries are now investing in quantum education or outreach for primary and secondary students~\cite{ISED2022,DISR2023,SQIS2022,M2025,NWGQT2021}, with Hong Kong having already integrated quantum computing into the high school curriculum~\cite{SZCFCLHZ2024}.

Two main issues in quantum computing education are the diversity gap (see~\cite{UF2025}) and the many prerequisites required by a conventional course in quantum computing~\cite{TGRP2022,ERB2020,IBBBDHKKLMPS2023,ASM2020,MM2024,LF2023,SFSL2021,SZCFCLHZ2024,PFKRM2020}.
In this paper, we describe the design and implementation of a two day outreach event for high school students, with a focus on recruiting underrepresented high school students in Nova Scotia, Canada.
The event is based on pedagogical principles such as mastery-learning, specification grading, and computer science education.
This event was organized as part of Canada's International Year of Quantum (see~\cite{CAP2025}).

The rest of this paper is structured as follows.
Sec.~\ref{Sect:Background} provides a review of relevant pedagogical research.
Sec.~\ref{Sect:Simulators} provides an evaluation of quantum circuit simulators for education.
Sec.~\ref{Sect:Design} describes the design of our curriculum and Sec.~\ref{Sect:Impl} describes its implementation.
Sec.~\ref{Sect:Eval} provides a quantitative evaluation of our program and Sec.~\ref{Sect:Lessons} discusses lessons learned.
\ifthenelse{\boolean{isfull}}{}{All appendices are found in the full paper~\cite{BDLW2026}.}

\section{Background and Motivation}
\label{Sect:Background}

\subsection{SMART Goals and Objectives}

Goal and objective setting are essential in both organizational management~\cite{O2017} and effective education~\cite{LH2012,WP2022}.
A goal typically refers to a long-term change in an indicator, whereas objectives are short-term tasks intended to achieve an immediate result~\cite{D1981,O2017}.
Due to the importance of goal setting, many heuristics have been developed to characterize effective goals (see,~e.g.,~\cite{O2017}).
In~\cite{D1981}, Doran introduced the SMART goal framework for organizational management which posits that effective goals should be \textbf{specific}, \textbf{measurable}, \textbf{assignable}, \textbf{realistic}, and \textbf{time-bounded}.
In individual goal-setting, a goal being assignable means that it is personally attainable~\cite{LH2012}.
In prior work, SMART goals have proven effective in an educational setting~\cite{LH2012,WP2022,MMCNR2024}.
For example,~\cite{WP2022} notes that after introducing SMART goals to a Management Fundamentals class, students reported starting projects earlier and enjoying their projects more.
Some critics of SMART goals argue that they overlook self-determinacy and the emotional dimensions of goal-setting~\cite{DT2011}.

More recently~\cite{O2017}, Ogbeiwi introduced a framework for writing SMART goals.
This framework is situated within goal-attainment theory, and notes that to write a SMART goal it suffices to write a single OITT sentence which includes an \textbf{outcome}, an \textbf{indicator} of success, a desired \textbf{target} level, and a realistic \textbf{time-frame} in which the goal must be achieved.

\subsection{Bloom's Mastery Learning}

Mastery learning was proposed by Bloom in 1968, after observing that students could master most topics provided adequate tutoring~\cite{G2007}.
Based on this, Bloom proposed a model of teaching in which educators would adjust their lessons to the needs of their students, so that students with high aptitude could take part in enrichment activities while students with lower aptitude could receive additional support~\cite{B1968}.
In a sense, Bloom accepted the claim that aptitude is normally distributed but rejected the claim that this would lead to a normal distribution of student outcomes.
Instead, Bloom claims that given adequate time and support, $95\%$ of students should be able to achieve an A-level grade.
In practice, this would never happen, since all courses are time-bound, and it is only natural that some students will require more time than is allotted by the course structure~\cite{SBSM2025}.
Despite this limitation, a recent meta-analysis of $108$ mastery-based courses demonstrated a grade increase of $0.52$ standard deviations on average~\cite{KKB1990}.

While there are many ways to implement mastery learning, the structure is always cyclic in nature~\cite{G2007}.
First, the teacher provides a lesson, after which students receive an ungraded assessment.
This assessment is used to differentiate students who have mastered the lesson from those who require more time.
The students who have mastered the lesson are provided with enrichment activities, while the struggling students are provided corrective activities to reteach the content through a new medium or perspective.
In either case, the choice of activity should target the aptitudes and interests of the student.
Following the activity, the class will reconvene, and the next lesson will commence.
A cycle condensed to a single lecture is referred to as a \emph{microcycle}~\cite{WP2022}.
In~\cite{WP2022} it is shown how these cycles align with self-determination theory, which explains the anecdotal phenomenon that students are more likely to engage with their classes when mastery learning is used.

The benefits of mastery learning are not limited to higher graders and engagement.
For example, students report lower anxiety~\cite{SBSM2025,HH2020,F2021}, greater self-efficacy~\cite{CSD2012,H2014}, improved outlooks on their courses~\cite{G2007}, a greater sense of self-determination~\cite{HH2020}, and greater self-confidence~\cite{G2007,F2021}.
Moreover, researchers have noted that mastery learning promotes a growth mindset~\cite{HH2020,F2021}, which has been shown to predictive of success in STEM courses~\cite{YHW2019,BTD2007}.
In terms of outreach, it has also been noted that these improvements to pedagogical practices may help to improve recruitment and retention of underrepresented groups~\cite{G2007,TH2022}.
Of course, mastery learning is not without its own challenges.
For example, students occasionally report higher levels of procrastination~\cite{SBSM2025,HH2020}.
Moreover, students sometimes have trouble adjusting to mastery learning, such as mistaking corrective activities for punishments~\cite{R2013} and misunderstanding how the new assessment schemes will work~\cite{HML2021}. 

\subsection{The Integrated Course Design (ICD) Framework}

\emph{Integrated Course Design (ICD)} is a framework for constructing courses in which the learning goals, assessments, and classroom activities are all aligned and informed by the situational factors unique to the course~\cite{F2013}.
The framework is based upon backwards design, which has proven effective in secondary education~\cite{WM2005}.
In this approach, the instructor first comes up with the learning goals, from which assessments are designed.
Once the assessments have been designed, the instructor can then work backwards to identify the lectures and activities required to prepare students for the assessments.
At each stage of course development, the alignment of the new components are assessed.
This framework has nine steps, not including revision to the course (see~\cite{F2013} for details).
\begin{enumerate}
\item Identifying significant factors unique to the course. 
\item Identifying important learning goals. 
\item Formulating appropriate feedback and assessments. 
\item Selecting effective teaching and learning activities. 
\item Ensuring the primary components are integrated.
\item Creating a thematic structure for the course. 
\item Selecting a teaching strategy. 
\item Integrating the course structure and the instructional strategy to create an overall learning activity scheme. 
\item Developing the grading scheme. 
\end{enumerate}

\subsection{Grading for Competency}

Mastery learning has gained recent popularity in STEM courses~\cite{Berns2020,CCO2020,C2020,HML2021,TAE2019,TH2022,P2020,F2021}.
One unique challenge of mastery learning is that of grading.
In a traditional grading system, students may receive part marks for incorrect work.
In principle, a student could pass a course entirely on part marks (see~\cite{SBSM2025}).
In contrast, the philosophy of mastery learning would suggest that students should only receive credit for their work once they have achieved mastery~\cite{B1968,WP2022}.
This line of inquiry has led to many alternative grading schemes, collectively known as \emph{grading for competency}.  

In theory, grading for competency can be divided into three main categories~\cite{G2016,CCO2020}: mastery-based, standards-based, and specification-based.
In \emph{mastery-based grading}, students are provided a list of outcomes for which they must demonstrate mastery by the end of the course.
Grades are determined by the number of outcomes mastered across multiple examinations.
When evaluation extends beyond examinations, this is called \emph{standards-based grading}.
A variation on this is \emph{specification grading}~\cite{N2014}, in which each assessment is accompanied by a detailed specification (e.g., a rubric), against which each outcome is graded on a pass/fail basis.
Essentially, the rubric describes all the ways in which a student could demonstrate mastery while completing the activity, and points are awarded only if mastery is achieved.
In all three variations, certain outcomes can be designated as \emph{essential}, for which failure to master the outcome constitutes failing the class~\cite{TAE2019}.

Specification grading is well suited for STEM courses, since even if a question is subjective, the identification of a proficient answer is typically objective~\cite{TAE2019}.
Occasionally, students will complain about the loss of part marks under specification grading~\cite{P2020,C2020,HML2021}, though this is usually offset by the ability to re-attempt any specification, with many students finishing the course with a positive perception on specification grading~\cite{P2020,C2020,HML2021,Berns2020,TH2022,F2021,TAE2019}.
It should be noted that simply adopting specification grading does not improve student performance~\cite{C2011}.
However, it is hard to integrate specification grading without embracing the principles of mastery learning, which accounts for the increase in achievement.
This can be reinforced by following a course design framework which embraces backwards design.
Best practices suggest that not only should backwards design be embraced, but moreover, all outcomes should be SMART~\cite{WP2022}.

\subsection{Hackathons as an Educational Tool}

A hackathon is a short (2--5 day) event, in which participants form teams and work collaboratively to solve programming problems, typically for prizes or recognition~\cite{PKIHKH2019}.
This model has been widely used in computer science as a form of project-based learning~\cite{PKIHKH2019}.
Due to the rigid structure of high-school curricula, hackathons have also been used as an extracurricular tool to introduce high-school students to topics from quantum computing~\cite{PFKRM2020}.
Quantum hackathons can be seen as a special case of the interdisciplinary hackathon, in which participants work to solve problems in domains such as health care~\cite{TPG2021,PSMMPDWC2017}, smart city design~\cite{AAGL2016}, and architectural design~\cite{SLLCG2015}.
It is argued in~\cite{TPG2021} that the interdisciplinary approach improves equity, since no single participant is an expert on all interdisciplinary knowledge.

Unfortunately, the interdisciplinary approach is insufficient to ensure participation from underrepresented groups~\cite{TPG2021}.
For example,~\cite{DEV2015} found that women were often hesitant to attend hackations due to negative preconceptions, such as the assumption that hackathons would be hostile, highly competitive, male-dominated environments.
Removing prizes and prioritizing beginners has been shown to improve the attendance of women at hackathons~\cite{DEV2015}.
In~\cite{SLLCG2015}, it was also suggested to avoid the use of the word \emph{hackathon}, favouring words such as \emph{designathon}.
Similarly,~\cite{DEV2015} notes that \emph{code jams} are perceived as friendlier than \emph{hackathons}.

\subsection{Cooperative Learning and Hackathons}

Cooperative learning describes any learning context in which students must work together to achieve their learning goals~\cite{FB2007}.
This could be as simple as encouraging students to collaborate on practice problems, or as complex as requiring students to complete a group project which spans the entirety of the course.
A recent meta-analysis has demonstrated that cooperative learning improves students’ achievement, especially when used within STEM classes~\cite{O2023}. 

One challenge to cooperative learning is \emph{norm-based grading}, in which the final grade of each student is obtained by fitting the grades of all students to a normal distribution~\cite{S2005}.
In this way, a students’ success is dependent on how well they can outperform other students in the class.
It has been demonstrated in multiple studies that when success in a class is linked to competition, the amount of collaboration between students is greatly diminished~\cite{FB2007} as predicted by Bloom~\cite{B1968}.

A recent model for cooperative education is that of the hackathon.
Since hackathons are group-based, this inherently encourages collaboration between small groups of students.
However, the structure of a hackathon is competitive in nature, which may inhibit collaboration between groups.
In practice, participants often work on personalized goals, and collaboration between teams still takes place~\cite{PKIHKH2019}.
In~\cite{PSMMM2022}, each team was provided a different problem to work on, to ensure that each team would have a unique sense of purpose.

\subsection{Computational Thinking (CT) and CT Tools}

In 1996, Papert~\cite{P1996} considered a world where K-12 mathematics was taught through programming, allowing students to interact directly with the mathematical objects of study.
In this world, students would learn to think of mathematics from an algorithmic point-of-view.
Along the way, students would build computer fluency, become better problem solvers, and master skills such as encapsulation and abstraction.
This educational journey would be aided by specially designed visual programming languages with dynamic components.

This notion was revisited by Wing~\cite{W2006} in 2006 under the name of \emph{computational thinking (CT)}.
At a high-level, CT encompasses the problem-solving processes utilized by computer scientists, such as algorithm design, abstraction, encapsulation, and simulation, all with respect to some model of computation.
In the same vein as Papert, Wing argued that CT is not limited to computer science~\cite{W2006}, and is applicable to other fields such as statistics, physics, and engineering~\cite{W2008}.
Later authors would extend CT to include concepts such as testing, debugging, and understanding algorithm design as a creative human process~\cite{BS2011,GP2013}.
Due to the diverse nature of computer science, it is not possible to list all processes involved in CT~\cite{EL2022}.
While there is some debate around the scope~\cite{BS2011,GP2013,D2017} and applicability of~\cite{D2017} of CT, it is clearly essential in a computer science education.
While it is still unclear how best to teach CT, researchers have suggested that symbolic abstractions~\cite{W2008}, the use of metaphors~\cite{AG2020}, scaffolding~\cite{AG2020}, and designing activities with gender equity in mind~\cite{GP2013,AG2020} are all important steps.

As suggested by Papert, a successful CT curriculum also relies on high-quality CT tools~\cite{RBE2016}.
These are visual programming languages in which students can formulate problems, express solutions, and evaluate the results in an iterative feedback loop.
Ideally, CT tools would provide cognitive frameworks for coming up with appropriate abstractions while avoiding accidental complexity in the overall solution.
Many approaches have been explored for the design of such tools, though no clear consensus has been reached on the best methodology.

\subsection{Block-based Programming Languages}

A block-based programming language is a language in which programs are built from nodes in an abstract syntax tree, with each node being represented as a puzzle piece~\cite{WW2015}.
The shape of each puzzle piece is used to indicate which syntactic constructions are valid, removing the possibility of syntax errors~\cite{WW2017}.
Each puzzle piece will include a natural language description to communicate to the end-user its intended purpose~\cite{WW2015}.
Typically, the blocks will also be colour-coded to reinforce to the end-user what each block is used for~\cite{WW2015}.
Puzzle pieces are joined together in a drag-and-drop fashion, with visual feedback indicating valid constructions~\cite{WW2015}. 

Block-based languages offer many advantages to novice programmers and are common CT tools~\cite{GP2013}.
For example, they eliminate syntax errors~\cite{WW2017}, reduce memorization~\cite{WW2015}, emphasize composability~\cite{WW2015,WW2017}, and improve readability~\cite{WW2015}.
In~\cite{WW2017}, it was shown that using such languages in introductory high school courses could increase grades, improve program comprehension, boost confidence, and encourage students to pursue computer science careers.
However, both~\cite{WW2015} and~\cite{WW2017} found that students could perceive block-based languages as ``less authentic'' and not representative of ``real programming''.
In contrast,~\cite{WJCW2013} found that when students could make real applications, this sense of inauthenticity was lessened.
Some research does suggest that students can feel discouraged when transitioning from block-based to plain-text languages~\cite{MLD2018}.
Despite this setback, test scores show that block-based programming skills are transferrable to a plain-text environment~\cite{WW2017}.
The results of~\cite{WW2015} and~\cite{WJCW2013} further show that if the block-based programming environment can convert each block into a plain-text representation, then the sense of discouragement is lessened.

In~\cite{KM2016}, heuristics were introduced to compare such novice programming systems.
In particular, the authors identified thirteen characteristics all novice programming systems should have: engagement, non-threatening design, minimal language redundancy, consistency, learner-appropriate abstractions, visibility of the program status, support for secondary notations (e.g.,~macros), clarity/simplicity, human-centered syntax (and terminology), edit-order freedom (e.g., the ability to leave some fragments incomplete), minimal viscosity (i.e. making changes should be easy), error avoidance, and feedback.

\subsection{Quantum Computing and Underrepresented Students}

Many quantum computing educational programs have been developed for K-12 students~\cite{ASM2020,PAHLC2020,ERB2020,SFSL2021,HITPS2022,MY2022,PSMMM2022,IBBBDHKKLMPS2023,SZCFCLHZ2024,SHRA2025,CPYWPCPWSTAKGKC2025,TFBLEW2019}, of which several consider outreach to underrepresented students~\cite{PAHLC2020,MY2022,IBBBDHKKLMPS2023,SHRA2025}.
By targeting outreach events at high school students, it is possible to reach underrepresented students before they make major career decisions~\cite{IBBBDHKKLMPS2023}, while also circumventing the challenge of modifying high school curricula~\cite{PFKRM2020}.
To reach underrepresented students coming from economically disadvantage social groups, it is also recommended to eliminate all costs associated with participating in the outreach program~\cite{IBBBDHKKLMPS2023}.
Of course, economic barriers are not the only factor limiting engagement from underrepresented groups.
For example,~\cite{RM2023} found that women were discouraged from completing computer science courses due to existing stereotypes and intimation in male-dominated settings.
This could be offset by providing these students the opportunity to engage with instructors who are women, and empowering them to use computer science for social good~\cite{RM2023}.
This is in line with the more general sentiment that representation and empowering students to apply what they learn to social issues can improve STEM engagement from underrepresented groups~\cite{GMGB2020,D2011}.

\section{Circuit Simulators in Quantum Education}
\label{Sect:Simulators}

Programs can be represented in many ways, such as by plain-text, abstract syntax trees, or control-flow diagrams.
A common representation in both classical reversible computing and quantum computing is that of the circuit diagram.
In this representation, each ``bit'' is represented by a wire which is acted on by gates.
These diagrams visualize many important program properties, such as causality and locality, which are hard to infer from plain-text.
However, circuit-based representations are not widely used in computer science courses, since neither classical reversible computing nor quantum computing are mainstream topics in a computer science curriculum.
This means that there is little research on the pedagogical implications of choosing circuit-based CT tools when introducing novice programmers to quantum computing.
On the other hand, CT tools such as graphical circuit simulators are already widespread in high school outreach, with 8 of the 12 educational papers surveyed making use of Qiskit Composer~\footnote{\url{https://quantum.cloud.ibm.com/}}.

However, these case studies and experience reports fail to provide a basis upon which quantum circuit-based CT tools should be chosen over their plain-text counterparts.
Moreover, Qiskit Composer is not the only quantum circuit-based CT tool.
For example, Quirk is a circuit-based CT tool for quantum program simulation\footnote{\url{https://algassert.com/quirk}}.
In this respect, the prior case studies and experience reports also fail to provide a basis upon which to compare Qiskit Composer to Quirk.
To overcome these two limitations, we turn to the wider body of literature on graphical programming languages in introductory programming courses.
We consider block-based languages, which share some similarities to circuit-based CT tools.

Based on this research, the use of circuit-based CT tools in an introductory quantum computing course does seem promising.
Many of the defining characteristics of block-based languages can be found in circuit-based languages.
Firstly, circuit-based languages expose the underlying syntactic structures hidden by plain-text quantum programming languages.
Moreover, circuit-based languages allow end-users to build circuits by dragging and dropping gates, whose shapes and labeling indicate their functionality.
However, there are no standard practices for labeling and colour coding gates.
Despite this, circuit-based languages still manage to eliminate syntax errors, reduce memorization through gate libraries, emphasize composability, and are often believed to be easier to read.
To compare Qiskit Composer and Quirk as educational tools, we therefore turn to best practices on the design of CT tools and the heuristic framework of~\cite{KM2016} (see Sec.~\ref{Sect:Background}).

\subsection{Comparison as CT Tools}

Neither tool was designed with an iterative feedback loop in mind.
However, both tools offer a live coding environment in which users can edit the circuit and see real-time changes in the simulated output.
In principle, a student could use this feature to iterate between problem formulation, solution implementation, and evaluation, though the student is not aided by the tool in this process.
Moreover, neither tool offers meaningful support for user-defined abstractions.
In this sense, both Qiskit Composer and Quirk are unsatisfactory CT tools.

\subsection{Comparison as Block-Based Programming Environments}

Due to their simplicity, both tools achieve consistency and minimal language redundancy in a trivial way.
Likewise, neither tool offers any meaningful support for secondary notation, edit-order freedom, or user-centric feedback.
By design, both tools visualize all dependencies and prevent users from making non-logical errors.
In terms of clarity, both tools are relatively comparable, and a user-study would be necessary to draw any meaningful distinctions.
For these reasons, we compare Qiskit Composer and Quirk only on the dimensions of engagement, non-threatening design, learner-appropriate abstractions, human-centered syntax, and minimal viscosity. 

\subsubsection{Engagement}
Neither Qiskit Composer nor Quirk are designed with student engagement in mind.
However, both tools offer unique ways to engage students.
In the case of Qiskit Composer, it is the ability for students to run their circuits on real quantum hardware at the click of a button.
This would give the students the sense that they really are \emph{doing} quantum computing.
In contrast, Quirk provides students with real-time feedback through the \emph{spinning} gates, which can be added to a circuit to visually simulate how the outputs of the circuit will change over time, given that the inputs are also changing over time.
Moreover, Quirk offers high-level gates such as \emph{frequency} gates (e.g., the quantum Fourier transform) and \emph{arithmetic} gates (e.g., adders and multipliers) which can enable students to design meaningful quantum circuits without needing to first master Fourier analysis and reversible logic. 

\subsubsection{Non-Threatening Design}
A non-threatening design is characterized by not overwhelming the end-user, building on easy to learn mechanisms such a drag-and-drop system, and making it easy for the end-user to undo their mistakes~\cite{KM2016}.
Clearly, both Qiskit Composer and Quirk build upon the same underlying mechanisms, which we believe to be relatively easy to learn.
Moreover, both Qiskit Composer and Quirk provide undo buttons with support for the \texttt{Ctrl+Z} key-binding used in most text editors.
However, Quirk is likely to appear far more overwhelming than Qiskit Composer to a first-time user.
This is because Quirk exposes all menus at once, without the option to suppress any panels.
For a beginner user, only 14 out of the 109 features would be necessary.
This level of complexity goes beyond visually distractions, and even proved confusing for the authors of this paper.
We conclude that Quirk has a relatively threatening design, when compared to Qiskit Composer. 

\subsubsection{Learner-Appropriate Abstractions}
The main abstractions used in Qiskit Composer and Quirk are for measurement and classical control.
In \appcite{App:Abstractions}, we describe how the abstractions used in Qiskit are less intuitive and more challenging for students without a background in physics.
We conclude that Quirk has abstractions which are more appropriate for quantum computing learners.

\begin{figure}[t]
    \centering
    \includegraphics[scale=0.4]{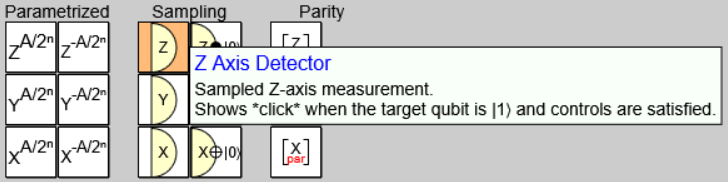}
    \caption{An example of unclear documentation in Quirk.}
    \label{fig:human-centered}
    \vspace{-1em}
\end{figure}

\subsubsection{Human-Centered Syntax}
The gates in both Qiskit Composer and Quirk use names and notations which would only make sense to those already familiar with quantum computing.
In Qiskit Composer, hovering over a gate simply reveals the same name as what is written on the gate.
On the ohter hand, hovering over a gate in Quirk will reveal a piece of documentation including the full name of the gate, a natural language description, and a visual representation of what the gate will do.
While the documentation is far from perfect (see Fig.~\ref{fig:human-centered}), it is still likely to help students recall the functionality of gates they have seen before.
We conclude that the syntax in Quirk is more human-centered, though neither tool is ideal.

\subsubsection{Minimal Viscosity}
The two main functions in Qiskit Compose and Quirk are adding gates to circuits and adding controls to gates.
Additionally, an end-user may wish to add a new wire to a circuit or change the input state of a wire in the circuit.
In terms of adding new wires, Quirk is less viscous, though the distinction is very minor.
In terms of adding controls to gates, Quirk is far less viscous.
A detailed analysis can be found in \appcite{App:Syntax}.
We conclude that Quirk has a far less viscous interface than Qiskit Composer.

\subsubsection{Additional Features}
One additional factor is that Qiskit Composer has a panel which displays an equivalent OpenQASM 2 program.
Editing the OpenQASM 2 program will edit the circuit diagram in real-time.
As is the case for block-based languages, this would likely help to ease the transition from a graphical circuit-based language to a plain-text language once the student is ready.
Moreover, the OpenQASM 2 output generated by Qiskit Composer is of far more use to a grader than the XML output generated by Quirk. 

\subsubsection{Conclusion}
Based on these four heuristics, we conclude that Quirk is far more appropriate for an introductory course on quantum computing.
Quirk might seem more threatening to students at first, but once a student overcomes that initial reaction, they would hopefully benefit from its many advantages, such as learner-appropriate abstractions and a less viscous interface.
One disadvantage of Quirk is that it does not offer an easy way to execute its circuits on real hardware.  

\section{Hackathon Design and Methodology}
\label{Sect:Design}

This section describes the design of our hackathon, and the pedagogical decisions underlying its design.
This hackathon is based on the principles of mastery learning and specification-based grading, with the ICD framework used to guide the overall design.
We note that this section omits step 9 from the ICD framework.
This is because step 9 is concerned with fromal grading, and is not necessary for an extracurricular hackathon.
To position our program within the existing literature, we draw on the lessons of prior quantum computing events for high school students~\cite{CPYWPCPWSTAKGKC2025,IBBBDHKKLMPS2023,ASM2020,PSMMM2022,PAHLC2020,HITPS2022,ERB2020,MY2022,SHRA2025,SFSL2021,SZCFCLHZ2024,TFBLEW2019}.

\subsection{Identifying Situational Factors (Step 1)}

In the ICD framework, a \emph{situational factor} is a factor external to the course content which may impact the success of the course~\cite{F2013}.
This includes: the specific context of the teaching and learning situation; the expectations of external groups; the nature of the subject; the characteristics of the learners; and the characteristics of the teacher(s).
A full analysis of the situational factors can be found in \appcite{App:Factors}.
The main insights from this analysis have been summarized below.
\begin{itemize}
\item There will be 4-6 lessons, ranging from 1 to 1.5 hours.
\item The event must be accessible for online attendees.
\item We must target the event at non-STEM students.
\item The event content should include social aspects.
\item We should teach using quantum pictorialism~(see~\cite{DYPPWCKGC2023}).
\item Our outreach efforts should target rural students, women, African Nova Scotians, and indigenous students.
\item The event should teach quantum computing abstractly.
\item We should collect student's backgrounds and interests.
\end{itemize}

\subsection{Identifying Important Learning Goals (Step 2)}

The ICD framework encourages educators to write learning outcomes that describe what a student should get out of a course, as opposed to merely listing the topics covered in the course~\cite{F2013}.
To ensure a holistic learning experience, it is suggested that learning goals span the following six dimensions of significant learning: foundational knowledge; application; integration; the human dimension; caring; learning to learn.
Using techniques described in~\cite{F2013}, an analysis of learning goals was carried out in \appcite{App:Goals}, from which the following learning goals were obtained.

\begin{enumerate}
\item Understand single qubit operations are rotations.
\item Understand the quantum teleportation protocol.
\item Identify factors limiting large-scale quantum computing.
\item Use Quirk and the ZX-calculus to simulate circuits.
\item Design simple quantum circuits to meet a specifications.
\item Identify the similarities between quantum and classical (reversible) computing.
\item Identify the interactions between quantum computing with the social sciences and humanities.
\item Understand the quantum mechanical phenomenon are complicated, and that it is okay to be confused.
\item Work productively in small teams to solve problems.
\item Value collaboration and diversity in scientific research.
\end{enumerate}

\subsection{Designing Feedback and Assessment Procedures (Step 3)}

The theory of mastery-based learning tells us that each lesson should be followed by a formative assessment in which the instructor validates the understanding of each student in a low-stakes environment.
During these assessments, instructors walk around the room and provide iterative feedback to the students regarding their progress.
If a student is struggling with a given concept, then the instructor will suggest a simplification to the problem with the intention of guiding the student towards a full understanding.
Likewise, if a student is performing particularly well, then the instructors will suggest more advanced variations of the problem, based on the interests of the students.
In either case, the instructors will employ specification grading, meaning that the students will be provided with clear guidelines on what to do, against which they will be graded on a pass/fail basis.
Since the assessments are formative, no formal grades will be recorded, and the words ``pass'' and ``fail'' will not be used explicitly.

On the second day, the students will take part in the hackathon which can be thought of as an educative and summative assessment.
In the language of the ICD framework, an educative assessment is any assessment from which a student is intended to learn new insights from, whereas a summative assessment is merely a graded assessment in the language of mastery-based grading.
In particular, we want students to experience what it is like to work on real quantum computing problems, as mentioned in the learning outcomes.
Of course, providing students will an open-ended quantum programming problem is not tractable after only one day of lessons.
Instead, we take inspiration from~\cite{PAHLC2020}, which claims that problems such as solving quadratic unconstrained binary optimization (QUBO) on a quantum computer can be taught to high school students, provided that the problems are adequately decomposed into sizable chunks.

To this end, we presented the students with three modules to work through, dealing with different aspects of quantum computing.
Each module consisted of one or more challenges, which in turn consisted of several objectives.
Students were asked to complete at least one challenge from each module, as a breath requirement (this is a form of essential objective, as described in~\cite{TAE2019}).
To allow high-performing students to stretch their abilities, optional \emph{advanced} objectives were also included in the challenges.

To align with best practices in specification grading, each objective was stated using a single OITT sentence.
Explicit time-bounds were omitted, to avoid hurting the self-esteem of struggling students.
Instead, we relied upon the time constraints imposed by the hackathon itself, together with the shared understanding that advanced objectives would be harder and should therefore require more time.
By giving students many challenges to choose between, we hoped to emulate the self-determinacy felt by students when setting their own goals.

\begin{figure}[t]
    \centering
    \includegraphics[scale=0.5]{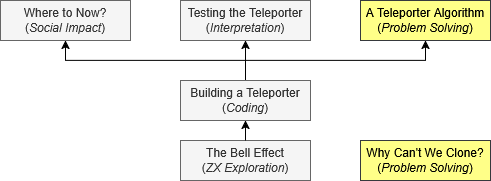}
    \caption{The objective structure for a teleportation challenge.}
    \label{figure:challenge}
    \vspace{-1em}
\end{figure}

To illustrate this approach, we will outline the assessment of a challenge which asked students to implement quantum teleportation.
As illustrated in Fig.~\ref{figure:challenge}, this challenge consisting of six objectives (the rectangles), two of which were marked as advanced challenges (the yellow rectangles).
The lines between rectangles indicate dependencies between objectives.
For example, a student must implement the quantum teleportation protocol before they can test the implementation.
When a student selects an objective, they are provided with a description of the challenge (based upon the OITT statement) and guidelines for what the student must submit.

\subsection{Designing Teaching and Learning Activities (Step 4)}

The ICD framework encourages a shift from passive to active learning, which aligns well with mastery-based learning.
Active learning includes doing activities, observing others performing activities, and reflecting upon activities.
While we will make use of passive lectures, we plan to compliment them with active content such as observing videos of the experiments we discuss, telling stories about the development of quantum computing (see~\cite{ERB2020,SFSL2021}), and visiting a laboratory working on quantum optics.
Through storytelling, we hope to give students a break from the more technical content while also working towards learning goals 8 and 10.
To improve the quality of our passive lectures, we plan to follow the advice of~\cite{MM2024} by including ample analogies.

As mentioned in the feedback and assessment section, each lecture will be followed by a formative assessment.
To make these assessments educative, we will have the students work through activities that apply what they learned during the lecture.
We took inspiration from quantum worm~\cite{PAHLC2020}, in the sense that the activities gave students puzzles to solve, but to solve the puzzles the students would need to apply what they had learned about quantum computing.

On the second day, students will have the opportunity to work on real-world problems.
This will allow the students to push their understanding, while also having a sense of authenticity in the work they are doing.
The challenges draw inspiration from the United Nations' Sustainable Development Goals (SDG's), to give the activities a sense of social impact.
After reviewing the ways in which quantum computing intersect with the SDG's, we settled on smart cities~\cite{SABK2024,BFFR2023}, which has been a successful theme for interdisciplinary hackathons in the past~\cite{AAGL2016}.
The competition was divided into three modules: quantum communication, quantum devices, and quantum optimization.
Challenges included exploring QAOA, implementing quantum teleportation and superdense coding, exploring quantum hardware, and learning how quantum sensing works.
The objectives making up each challenge drew on examples from the literature~\cite{PSMMM2022,ASM2020,M2022,DYPPWCKGC2023}, such as programming, hardware exploration, testing, ZX rewriting, and turning tasks such as oracle design into exercises.
By providing students with such an array of challenges, we hoped to simulate the unique sense of purpose described in~\cite{PSMMPDWC2017}.

\subsection{Integration of the Primary Components (Step 5)}

This step ensures that the learning goals align with the activities and assessments, and that the significant factors align with the design choices.
For most learning goals, the corresponding assessments are clear.
The formative assessments for learning goal 8 will follow from the impromptu discussions with mentors when students are struggling with the content.
There are currently no assessments for learning goal 10, and we plan to revise this in future iterations of the event.
In terms of alignment with significant factors, we have decided not to include hackathon in the title of event, to avoid the negative connotations associated with this word.
To emphasize the social aspects of this program, we have decided to name the event ``Quantum Hacking a Better Future''.

\subsection{Creating a Thematic Structure (Step 6)}

In the ICD framework, a \emph{thematic structure} is a list of four to seven essential topics, organized into a coherent sequence~\cite{F2013}.
We have identified the following six topics, which have been sequenced based on how the topics build upon one-another: (1) Qubits and single qubit operations; (2) Single qubit measurements; (3) Conditional statements and multi-quibt operations; (4) Quantum hardware and photonic systems; (5) The quantum teleportation protocol; (6) Applications in communication, devices, and optimization.
The thematic structure follows the recommendations of~\cite{MS2020}, which found that quantum outreach in a high school setting should choose a few topics to explore in full detail.

\subsection{Revisiting the Teaching Strategy (Steps 7 and 8)}

As suggested throughout the preceding steps, the teaching strategy employed by this event is a mix of mastery-based learning and problem-based learning.
The first day of the event is structured as a series of microcycles, in the sense of mastery-based learning.
This means that an instructor would first teach a concept to the students, and then the student would validate their understanding using the formative assessments described in steps 3 and 4.
Given that the event would only last two days, we concluded that the procrastination usually experienced during a mastery-based learning course would not be an issue for our students.
On the second day, elements of problem-based learning were also integrated into our teaching strategy, since hackathons are a form of problem-based learning.
Despite this shift in teaching strategy, our instructors were still expected to check in with students during the hackathon, and suggest remedial exercises when appropriate.
Since the event is only two days and was designed with these strategies in mind, it follows that the strategies are well integrated.

\section{Implementation and Participant Experience}
\label{Sect:Impl}

This outreach event was designed with the goal of empowering high-school students from under-represented groups in quantum computing community, especially those from rural communities and those who self-identify as women, non-binary, indigenous Nova Scotian, and African Nova Scotian.
To achieve this, we disseminated promotional material and an event description through established outreach organizations in the region such as KINU~\cite{Kinu}, Imhotep~\cite{Imhotep}, PLANS~\cite{Plans}, SuperNOVA~\cite{SuperNova}, and Prep Academy~\cite{Prep}.

The two-day event was held on August 8-9 in 2025.
The format was hybrid (using Zoom) to accommodate students from rural areas, with one student switching from in-person to online attendance.
The in-person students were provided with food (at no associated cost) to offset economic barriers.

\begin{figure}[t]
    \centering
    \includegraphics[scale=0.03]{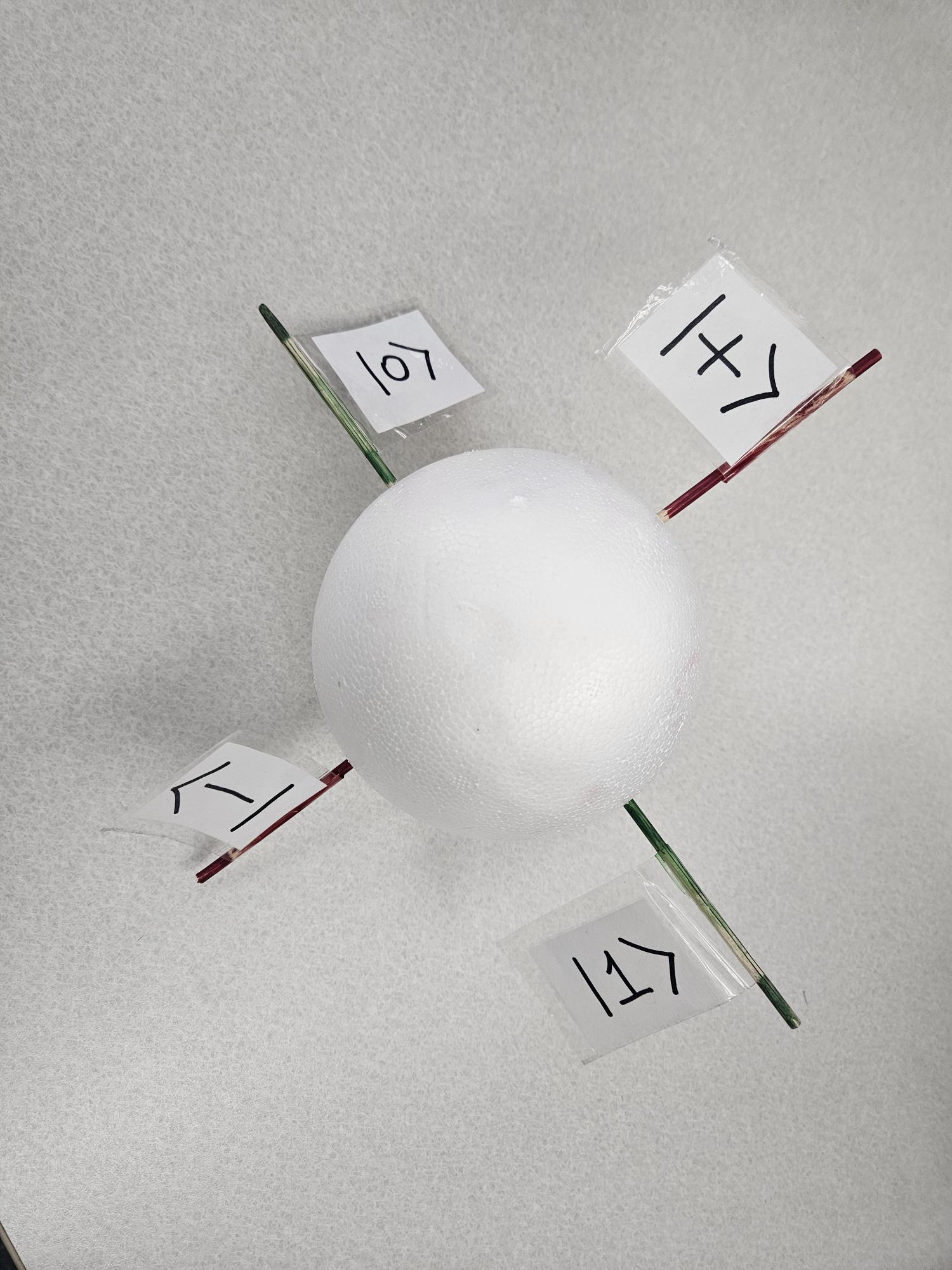}
    \caption{The physical Bloch Sphere.}
    \label{figure:bloch-sphere}
\end{figure}

The first day was divided into four lectures, separated integrated activities and breaks.
The introductory lecture focused on qubits, single-qubit operations, and their representation via ZX-calculus.
The lesson was designed to bridge abstract quantum theory with tangible physical analogies.
For example, the instructor used a light switch in the room as an analogy for classical bits and NOT gates.
To transition to quantum states, the instructor asked the students to consider the case in which the light switch is stuck between off and on (which was then used to explain the concept of superposition).
To help students grasp the symbolic abstraction of the Bloch sphere, we utilized a styrofoam ball equipped with axial labels for the x- and z-bases (see Fig.~\ref{figure:bloch-sphere}).
The instructor showed the physical Bloch sphere to the online participants, while letting the in-person participants manipulate the sphere to help complete activities.
Through this sphere, they could visualize the qubit vector states and the geometric effects of the single-qubit gates.
The three other lectures covered the topics of measurement, multi-qubit operations, and entanglement. 
At the end of the first day, participants identified the topic they found the most difficult, for review on the second day.

\begin{figure}[t]
    \centering
    \begin{subfigure}[t]{0.23\textwidth}
        \centering
        \includegraphics[scale=0.045]{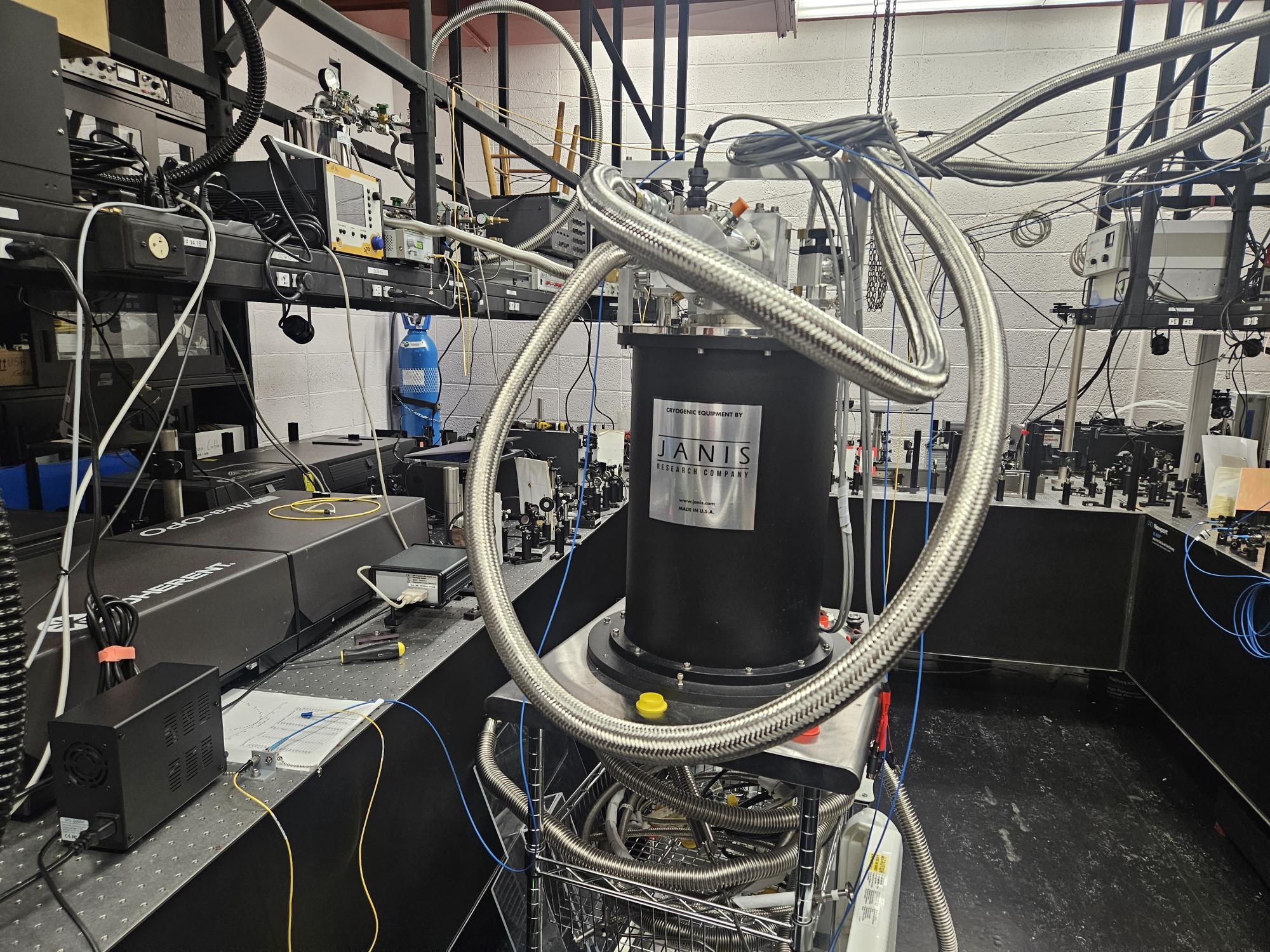}
        \caption{The cryostat in the lab.}
    \end{subfigure}
    \begin{subfigure}[t]{0.23\textwidth}
        \centering
        \includegraphics[scale=0.045]{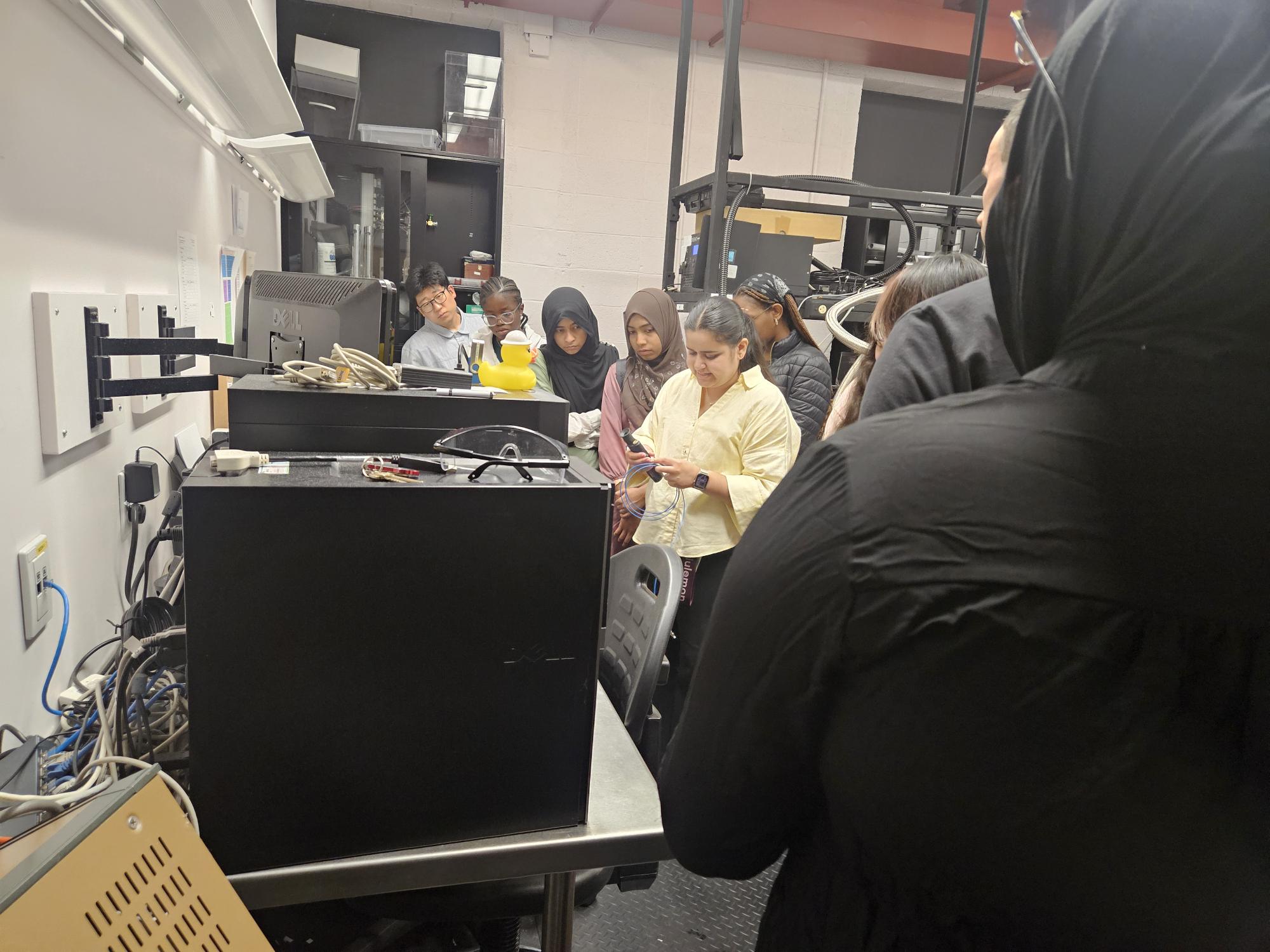}
        \caption{The lab instructor explaining photonic quantum computers.}
    \end{subfigure}
    \begin{subfigure}[t]{0.23\textwidth}
        \centering
        \includegraphics[scale=0.045]{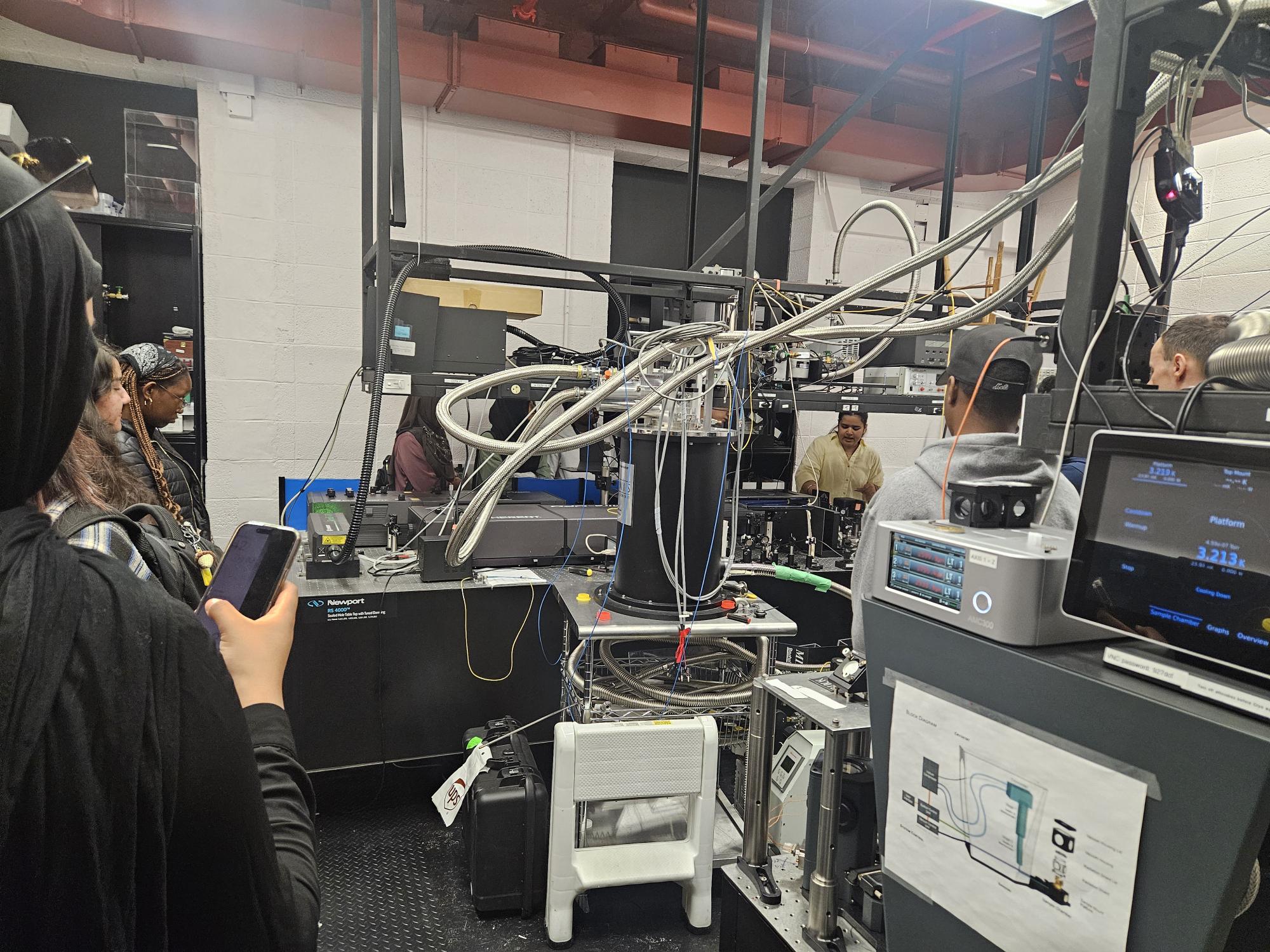}
        \caption{An experiment demonstration with lasers and mirrors.}
    \end{subfigure}
    \begin{subfigure}[t]{0.23\textwidth}
        \centering
        \includegraphics[scale=0.045]{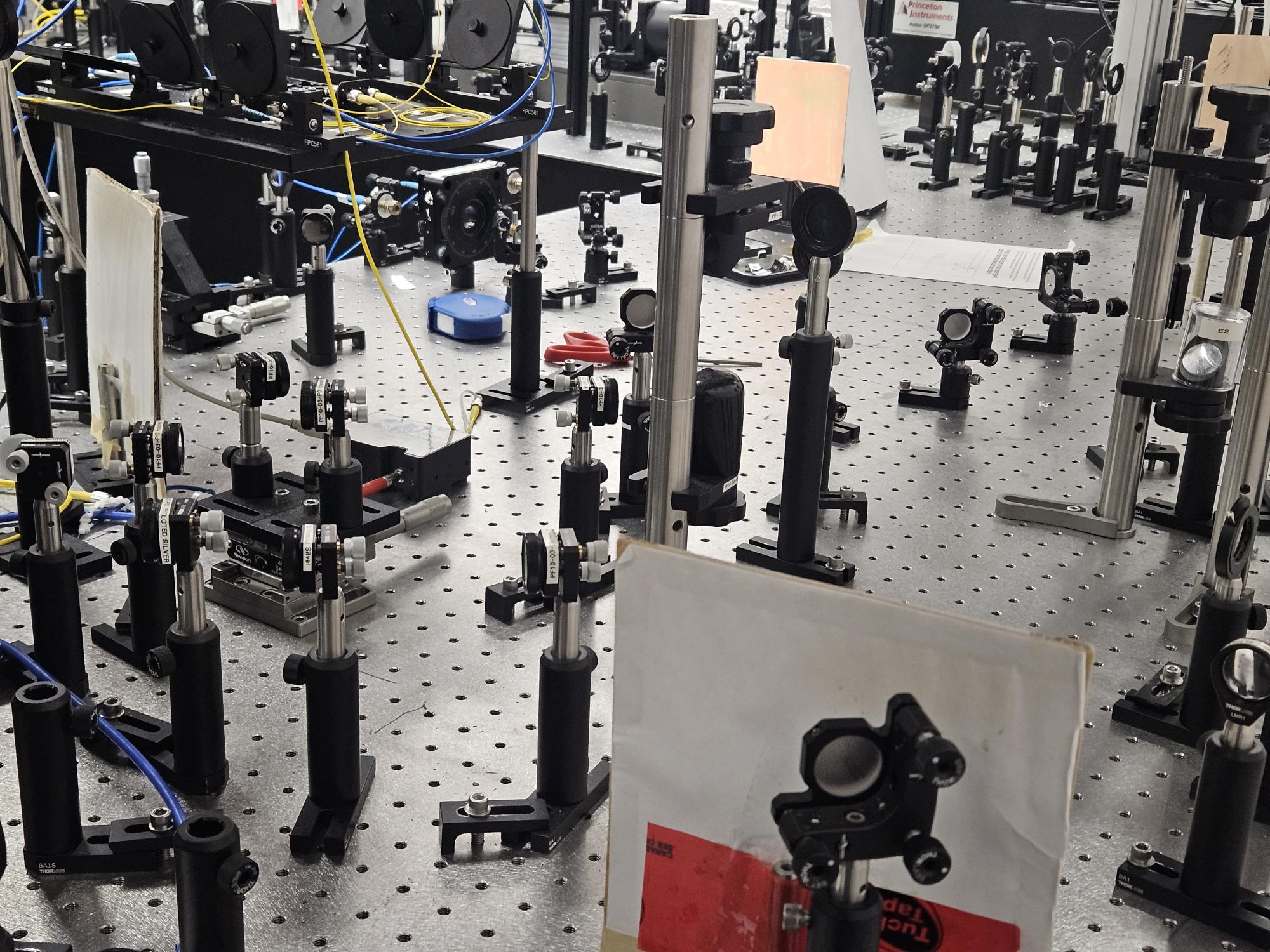}
        \caption{An optical table.}
    \end{subfigure}
    \caption{The visit to Ultrafast Quantum Control Lab.}
    \label{figure:lab-visit}
    \vspace{-1em}
\end{figure}

One highlight of the first day was a tour of the Ultrafast Quantum Control Lab at Dalhousie University (see Fig.~\ref{figure:lab-visit}).
During this session, the lab instructor guided the participants through quantum photonic experiments and demonstrated the principles of reflection using mirrors.
Also, the students tried their hands at assembling mirror stands for the experiments.

The second day of the event was allocated to the competition.
Before finalizing the challenges, the interests of the participants were taken into consideration.
For example, from the pre-event survey we know that one student was interested in anthropology and sociology, so one of the challenges we designed was for the students to analyze the social impacts of quantum computing.
For this challenge the students needed to use what they learned during the lectures about the advantages of quantum computing, together with the hardware limitations they had learned during the lab tour (e.g., the sub-zero conditions in which most quantum computers work).
Overall, the students were presented with a wide range of challenges including analyzing social impact of quantum computing, writing quantum programs, and using the ZX-calculus to predict the outcomes of quantum computations. 

\begin{figure}[t]
    \centering
    \begin{subfigure}{0.23\textwidth}
        \centering
        \includegraphics[scale=0.045]{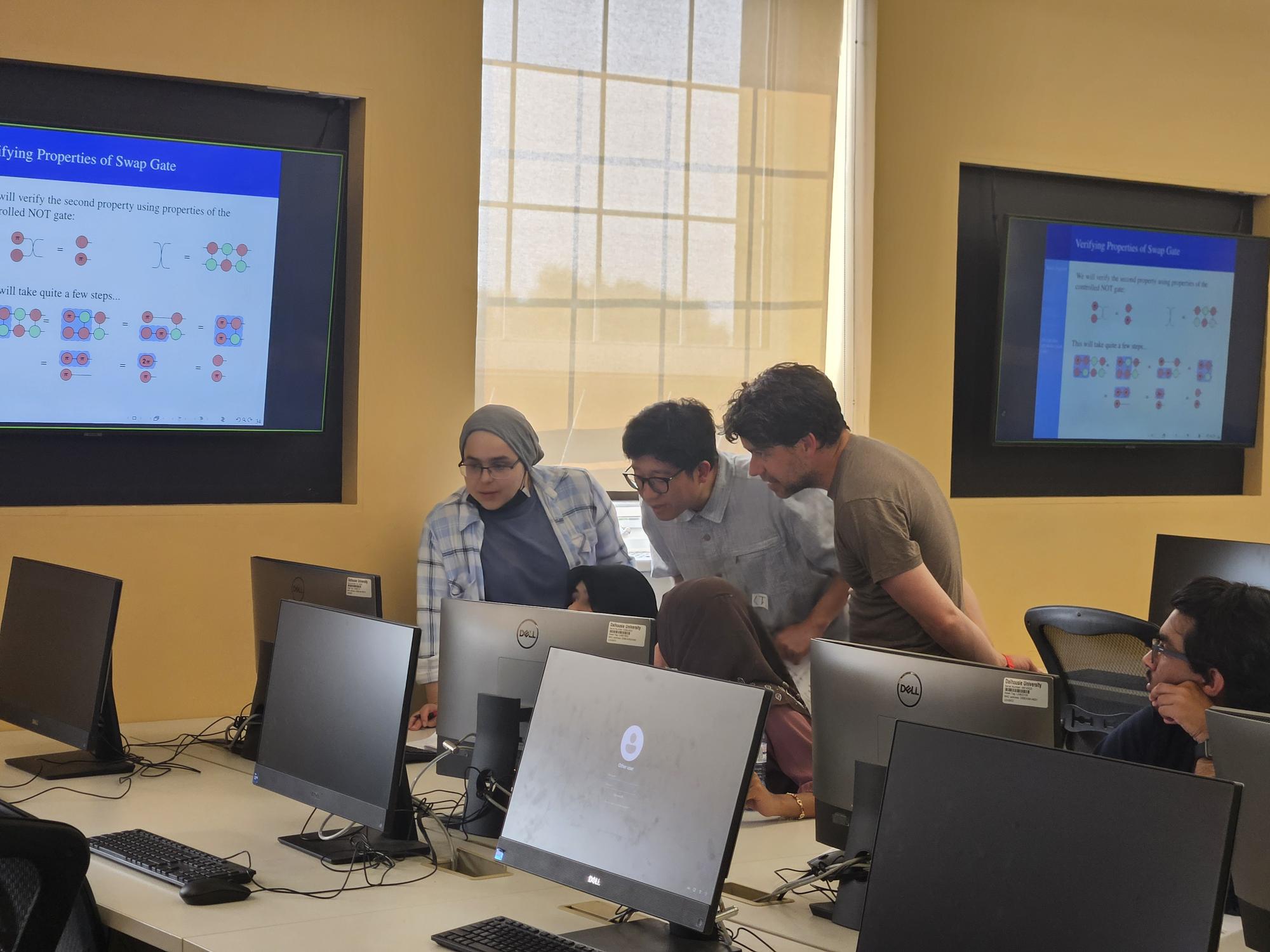}
    \end{subfigure}
    \begin{subfigure}{0.23\textwidth}
        \centering
        \includegraphics[scale=0.045]{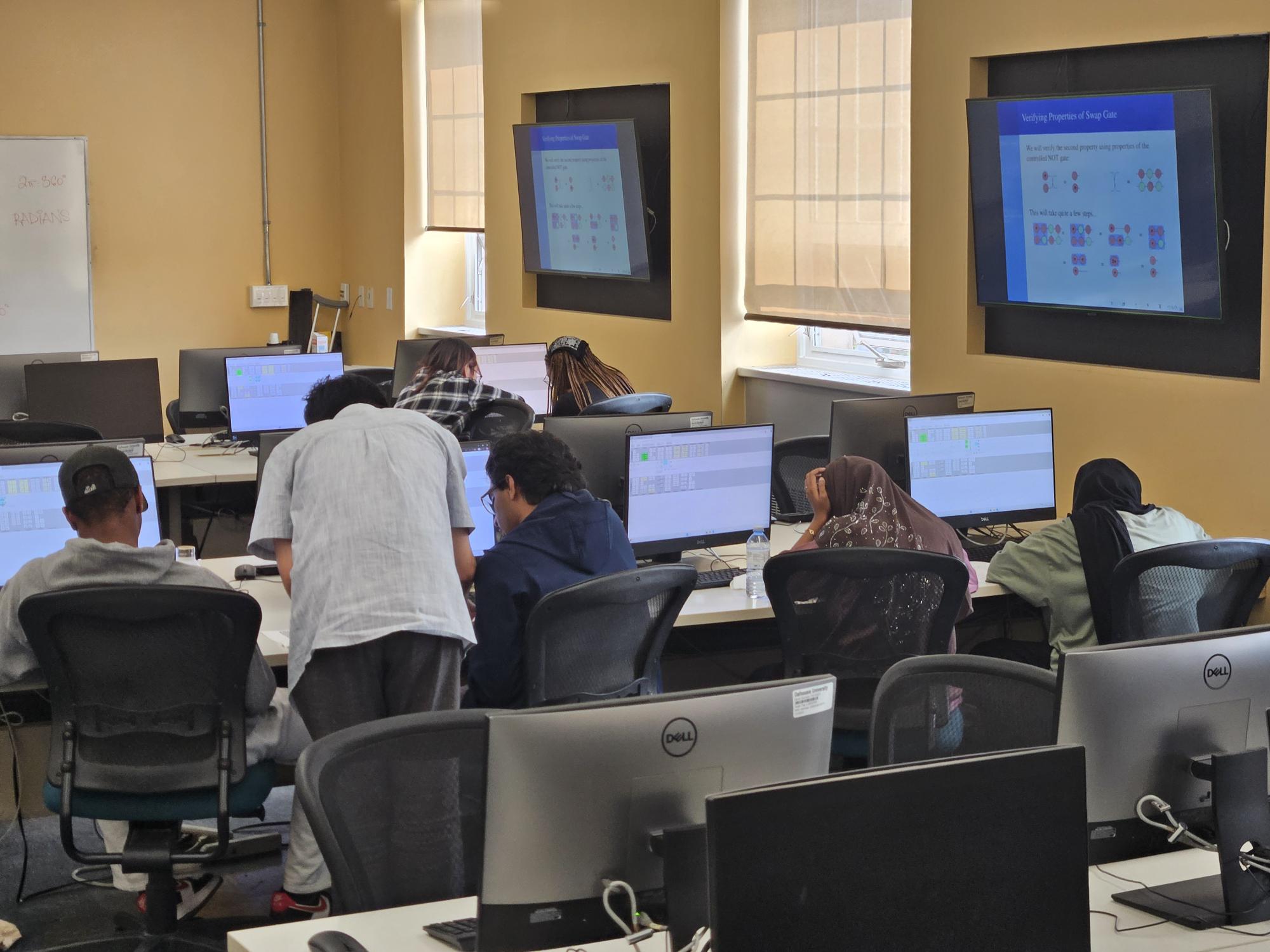}
    \end{subfigure}
    \caption{The second day of the event: The competition}
    \label{figure:the-competition}
    \vspace{-1em}
\end{figure}

During the competition, the instructors moved around the room to encourage students and answer their questions while they worked (see Fig.~\ref{figure:the-competition}).
We found that some students struggled to start working on challenges, due to limited self-efficacy.
The instructors tried to encourage these students, by emphasizing that the goal was the process of problem solving, rather than getting a perfect solution or winning the competition.
While evaluating the results of the competition, we found these students managed to solve several challenges, showing that not only had they gained the necessary understanding of the basic concepts, but also had moved towards a growth-mindset and as a result gained confidence in their abilities.

When we designed the challenges, we were worried that they would be too hard for the students.
However, our main goal for the second day was for the students to work on real-world problems, and also to gain confidence and a sense of accomplishment in quantum computing.
The wide range of challenges, as well as their different formats (e.g., qualitative written analysis, drawing ZX-diagrams, simulating circuits in Quirk, and coding) ensured that the students could engage with the topics they found interesting using the medium they felt confident in.
The type of the challenges the students chose was a good indication of both their interest and the area they felt most comfortable with.

\section{Outcomes and Evaluation}
\label{Sect:Eval}

Both before and after our event, we distributed surveys to the students to evaluate whether we reached our target demographics and whether our hackathon had an impact on the students' relation to quantum computing.
These surveys were approved by Dalhousie University's Research Ethics Board under the exemption for program evaluation.
The data collected includes demographics, educational backgrounds, impressions of quantum computing, feedback on the event, and a freeform response question about quantum computing.
In addition to these surveys, the performance of the students was also evaluated based on the number of objectives each team solved on the second day of the hackathon.
The raw data can be found in \appcite{App:Data}.

In total, $10$ students attended the event, with many of these students identifying as women and African Nova Scotians.
According to the survey results, the students become more familiar with the quantum computing topics covered in the event.
Unfortunately, only half of the $10$ students who attended the lectures on the first day participated in the competition on the second day ($3$ students reported conflicts).
Moreover, the performance of the students who participated in the competition varied greatly, with objectively completed ranging from $2$ to $10$.
Nevertheless, those who came to the second day, responded that they enjoyed the challenges and developed stronger opinions on quantum computing.
In particular, to a question asking students to write a few sentences explaining a topic from the event to a friend, the responses from $5$ students covered most of the topics highlighted in the event, such as qubits, superposition, entanglement, measurement, the advantages of quantum computers, and how they differ from classical computers.
From this data, we conclude that most of the students enjoyed the event and learned some aspects of quantum computing, although the level of learning varied greatly across the students.

The rest of this section goes through the demographic data, survey results, and the students performance in more detail.

\subsection{Demographic Data}

In total, $20$ students registered from which $10$ students attended.
Half of the registrants and $80\%$ of the participants identified as woman/nonbinary.
Moreover, $40\%$ of the registrants and participants identified as African Nova Scotian.
None of the students self-identified as indigenous.

We collected further demographic data from an optional pre-event survey in order to better understand the educational backgrounds of our students.
Of the $20$ students who registered, $10$ students completed these demographic questions.
We found that $80\%$ of the participants lived in Halifax or Dartmouth, which constitute the only major urban center in mainland Nova Scotia.
Moreover, $90\%$ of these students had parents who attended university.
The grade levels of these students at the time of registration were greatly dispersed, ranging from $8$ (entering high school) to $12$ (graduating high school).
We had attempted to collect data on relevant courses taken by these students, but failed to account for the IB program offered in Nova Scotia.
We did find that the non-IB students had never taken computer programming.

Evidently, we achieved our goal of recruiting girls, and failed to recruit indigenous students.
Since African Nova Scotians make up $0.4\%$ of Nova Scotia's population~\cite{Census}, we do feel that we reached our goal of recruiting African Nova Scotians.
On the other hand, rural students make up $39.1\%$ of Nova Scotia's population~\cite{Housing}, so we feel that we have failed to recruit effectively from this demographic group.

\subsection{Evaluation on the Event}

On both the pre- and post-event survey, students were asked to answer questions regarding their interest in quantum computing and STEM using a Likert scale (responses take values from $1$ to $5$ where $1$ stands for ``strongly disagree'', $3$ stands for ``neutral'', and $5$ stands for ``strongly agree'').
Students were also asked to identify familiar terms associated with quantum computing.
The surveys asked the same questions before and after the event to identify shifts in perspectives.

First, we asked whether students would like to attend university, to which we saw no shift before or after the event.
We then asked students whether they would like to work in a STEM-related field, to which we saw one student shift from ``neutral'' to ``strongly agree'', which all other answers remaining the same.
Finally, we asked if students would like to work with quantum computers, to which we saw two students shift from ``neutral'' to ``agree'' and one student shift from ``neutral'' to ``disagree''.
These shifts away from neutral suggest that our event helped students to develop their opinions on STEM and quantum computing.
Naturally, not all students are interested in quantum computing.

We also found that the number of quantum computing terms recognized by the students generally increased following the event.
To our surprise, the number of students who recognized ``quantum measurement'' did not increase from $1$, and the reason is not clear.
We found that one less student felt familiar with ``Schrodinger's Cat'' after the event, which may reflect that fact that we dispelled misconceptions about the thought experiment during the second day of the event, after which the student may have felt less certain about the term.

We also found that students generally enjoyed the first day of event.
On a Likert scale, students answered agreed on average to having enjoyed the event.
For each of the lectures, students also agreed on average to having enjoyed the content, though these numbers varied from $3.71$ to $4.42$.
The students also agreed, on average, to having enjoyed the hands-on activities which took place at the end of lectures.
This suggests that our microcycles were well-received.

For the competition, feedback was collected from all $5$ students who attended the second day.
They agreed, on average, to having enjoyed the challenges, but were neutral on average to feeling motivated to complete the challenges.
This suggests more work is needed to successfully motivate students.

\section{Discussion and Challenges}
\label{Sect:Lessons}

The event was a success, especially given it was the first time hosting such an event for the organizers.
As discussed below, there were certainly challenges met along the way, but registration numbers surpassed our expectation, and participants were engaged in the hackathon.
A significant proportion of students were in underrepresented groups we aimed to reach, and according to our post-event survey, we had managed to spark interest in a career in STEM for a few students.

We encountered three main challenges over the course of the event.
First, there was a low level of engagement from participants during the lectures.
Most students erred on the side of caution when it came to voicing their thoughts regarding new topics, especially around strangers.
Some of the younger students also explained that they were less inclined to ask questions, in fear that it was already clear to the older students.

Second, it was difficult promoting group work during the hackathon (Learning Goal 9), without being forceful.
The hackathon challenges were designed to accommodate both individuals and teams alike, though teamwork was encouraged to stimulate cooperative learning.
Despite the advantages of being in a team (e.g., more ideas, dividing the workload), there was only one team on the day of the challenge.
Furthermore, the team members were already friends from the same school.
The reason for this, most likely, would be due to the fact that participants only had one day of lectures and activities to get acquainted with one another before committing to a team.

Underlying the previous two challenges is the third obstacle: two days were not enough for the event.
The lecture component of the first day is a crucial prerequisite for the following hackathon; however, there was more material presented than time given to digest all the new information.
Participants might have benefited from more time practicing the lecture content, or just get to know each other better before team formation.

On the other hand, these challenges also offered important insights to us as organizers.
We see that we failed to find the right balance between instructional time and break time, where the latter could be used for less structured activities such as discussions on the social aspects of quantum computing (as in~\cite{SFSL2021,IBBBDHKKLMPS2023}), ice breaking activities to ease the participants in, or hands-on sessions for more practice.
The initial aversion to engagement might persist, but additional group activities would, at the very least, make students feel more comfortable amongst their peers.
The allotted one day was definitely not enough to accommodate these extra activities, so extending the event is necessary.
This, in turn, would also provide us more time to pace our lectures with more intermediary materials for a smoother introduction, such as discussing the quantum computer system stack (see~\cite{SFSL2021} and~\cite{FRLASVCB2016}).

Upon reviewing the lecture notes, we realized that while we provided students with the foundations of quantum computing, computational thinking was not explicitly emphasized.
We discussed how to view quantum mechanics in a computational way, and how circuit languages are built on layers of abstraction (e.g., the Bloch sphere, gates), but most of this was implicit.
There should have been more time spent discussing how one can apply abstractions in problem solving and formulate solutions algorithmically, which would have benefited the students greatly going into the challenges of the second day.
We found that students had familiarized themselves well with the lecture material, but faced difficulties when trying to translate this knowledge into challenge solutions.

In terms of future expansions, we need to consider both the scalability and sustainability of the event.
As we have noted above, we should lengthen the duration of the event.
To do this, we will need more preparation time for planning lectures and organizing activities (the two day even required 8 months).
Since most participants are minors, it is also necessary to find chaperones (potentially lecturers) to supervise the students.
Scalability, on the other hand, poses a dilemma.
On one hand, we could make this a yearly event, with the same skeleton for the lecture notes (modulo adjustments to student interests) while recruiting new junior and senior high school students.
Unfortunately, past participants may lose interest since they have already seen the material.
Alternatively, we could host intermediate-level hackathons in the following years to attract past participants and students who already have a grasp on the prerequisites.
However, this option runs counter to the goal of reaching students not previously interested in STEM.

Although the event is now over, there is still much for us to do in terms of final touches, as well as future improvements.
One of the lectures had involved a model Bloch sphere with the x- and z-axes label, which stirred much interest from participants; therefore, integrating more props could make the lectures more engaging and exciting for students.
In terms of our lecture notes, we would like to align the presentation more closely with computational thinking before releasing the contents.
In terms of the activities, we would like to explore integrating ZX tools such as \texttt{quantomatic}~\cite{KZ2015}, and recasting the hackathon objectives in terms of quest-based learning~\cite{S2022}.
Finally, our qualitative analysis in Sec.~\ref{Sect:Simulators} showed that both mainstream circuit simulators, Quirk and Qiskit Composer, have their drawbacks as educational tools.
Neither interfaces were beginner-friendly, especially since our event is targeted at students who might not have a previous interest in STEM.
The simulators did not have all the features we would like to demonstrate with the lecture contents, and the learning curve was steep if students wanted to attempt any advanced settings.
This will be something to study further--whether we compromise with the current simulators and find a way to go around obstacles, or we design a better option.

\section{Conclusion}

In this paper we described the design and implementation of a two-day quantum hackathon for underrepresented high school students in Nova Scotia, Canada.
The design of the event included an extensive literature review, and preliminary steps towards computational thinking tools in quantum computing education.
Based on our experience report and feedback collected from students, we conclude that this approach to quantum computing education is an effective way to introduce novice high school students to the key concepts of quantum computing.
In particular, we note that not only did students develop skills related to quantum computing, but also developed a sense of self-efficacy and a growth mindset.
In future work, we would like to address challenges such as encouraging engagement from students, promoting group work, and reaching rural communities in Nova Scotia.
We would also like to explore other graphical formalisms and the gamification of challenges.
Before releasing our lecture notes, we also plan to improve their alignment with CT and consider other CT tools such as \texttt{quantomatic}.

\section*{Acknowledgment}

We thank:
Neil J. Ross for his departmental support;
Sinmi Ayantoye for her event coordination and outreach support;
Andrew Allen and Lia Yeh for their curriculum feedback;
the Ultrafast Control Group and Jasleen Jagde for the lab visit.

\bibliographystyle{IEEEtran}
\bibliography{generic}

\newpage
\ifthenelse{\boolean{isfull}}{\appendix\subsection{Analysis of Learning Appropriate Abstractions}
\label{App:Abstractions}

This section compares the measurement abstractions, classical data abstractions, and classical control abstractions in both Qiskit Composer and Quirk.

\subsubsection{Measurement Abstractions}
In some sense, Qiskit Composer is a visual interface to the OpenQASM 2 programming language.
Then it should come as no surprise that Qiskit Composer adopts the same measurement abstraction as OpenQASM 2.
This means that each measurement is viewed as an experiment apparatus, which stores the measurement outcome to a pre-allocated classical bit, while simultaneously projecting the measured qubit onto the corresponding eigenspace.
This means that each measurement gate has a side-effect (storing the outcome to a classical bit) which is illustrated graphically via a dotted line from the measurement gate to a classical register.
In contrast, a measurement gate in Quirk simply converts the quantum wire into a classical wire which stores the measurement outcome.
The abstraction used in Quirk hides the physical details of how measurement devices work and aligns more closely with the conceptual model that a measurement will ``collapse'' a quantum state to a classical state.
This conceptual model, while not entirely accurate from a physical point of view, does align more closely with how students are first taught about measurement in quantum computing.
For this reason, the measurement abstraction used in Quirk seems to be more appropriate for novice programmers without a background in physics. 

\subsubsection{Classical Data Abstractions}
To understand how Qiskit Composer and Quirk differ in terms of classical control, it is necssary to first understand how they differ in terms of classical data.
In Qiskit Composer, a classical wire in a circuit represents an entire register, which may correspond to one or more classical bits.
For example, a single classical wire may encode a 16-bit integer.
On the other hand, each classical wire in Quirk corresponds to a single classical bit.
For advanced users, the Qiskit Composer abstraction has several benefits.
For example, if the first $n$ qubits in the circuit encode a binary input, then they can be read into an $n$-bit classical register which can then be interpreted as an $n$-bit unsigned integer rather than simply a bit string.
However, novice programmers are unlikely to think of such type cohersion when implementing their first quantum algorithms.
Instead, they are likely to think of each measurement outcome as a Boolean variable, which aligns with the abstraction used in Quirk.

\begin{figure}[t]
    \centering
    \includegraphics[scale=0.43]{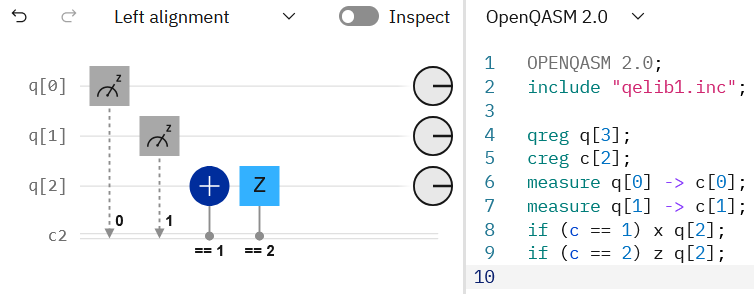}
    \caption{An example of control misuse in Qiskit Composer.}
    \label{figure:composer-ctrl}
    \vspace{1em}
\end{figure}

\begin{figure}[t]
    \centering
    \includegraphics[scale=0.4]{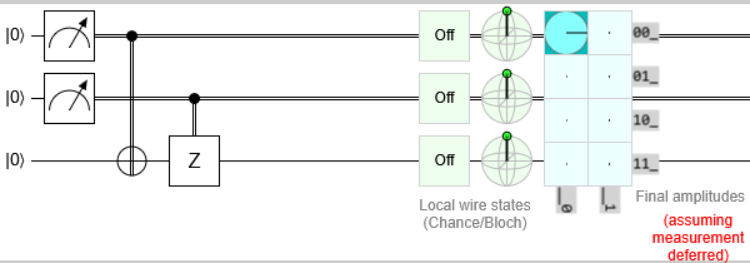}
    \caption{An example of control use in Quirk.}
    \label{figure:quirk-ctrl}
\end{figure}

\subsubsection{Classical Control Abstractions}
The complexity of the classical data abstraction used in Qiskit Composer carries over to the classical control abstraction.
The difficulty of working with this abstraction is best illustrated by the circuit in Fig.~\ref{figure:composer-ctrl}, which one author of this paper wrote (using the default measurement scheme) while trying to learn Qiskit Composer.
The circuit is meant to implement the corrections used in quantum teleportation.
That is, the $X$-gate is executed when the first measurement has outcome one, and the $Z$-gate is executed when the second measurement has outcome one.
However, the conditions $(==1)$ and $(==2)$ state that the $X$-gate should be executed when the bit string is $01$ and the $Y$-gate should be executed when the bit string is $10$.
Putting aside issues of endianness, which novice programmers ought not to think about, this code is still incorrect, since neither gate will execute when the measurement outcome is $11$.
The correct way to fix this bug would be to expand the classical wire into multiple wires, each containing a single bit.
Alternatively, an additional $X$-gate and $Y$ gate could be added to the circuit, each with the condition $(==3)$, though this solution is clearly not scalable.
Since classical wires in Quirk are necessary single qubits, then it is only possible for a novice program to write the collect implementation illustrated in Fig.~\ref{figure:quirk-ctrl}.
We conclude that the control abstraction used in Qiskit Composer is not appropriate for novice programmers, regardless of their background, whereas the control abstraction in Quirk aligns with student expectations.

\subsection{Analysis of Viscosity}
\label{App:Syntax}

This section compares the viscosity of Qiskit Composer and Quirk in terms of adding primitive gates (and wires) to circuits, adding controlled gates to circuits, and changing the input states of existing quantum wires.

\begin{figure}[t]
    \centering
    \includegraphics[scale=0.4]{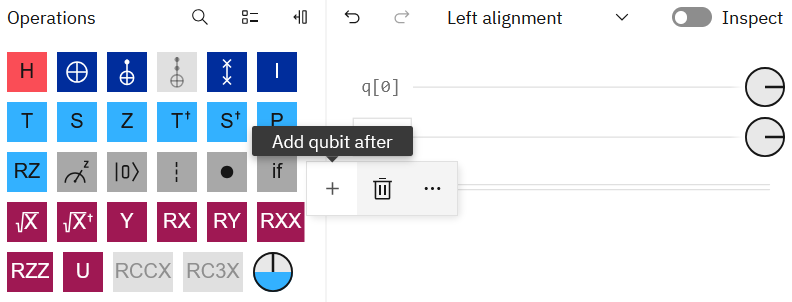}
    \caption{Adding a new wire in Qiskit Composer.}
    \label{figure:new-wire}
    \vspace{1em}
\end{figure}

\subsubsection{Viscosity and Adding Primitive Gates}
In terms of adding primitive gates, both Qiskit Composer and Quirk are relatively comparable.
All standard Clifford+$T$ gates are available from the interface and can be added to a circuit through a simple drag-and-drop interface.
Arguably, the Quirk interface is slightly less viscous, in that dragging a gate to the bottom of the circuit allows for the gate to be added to a new wire.
In Qiskit Composer, the same behaviour requires adding manually adding a new wire to the circuit.
This is then by first clicking on the name of a pre-existing qubit, and then selecting the plus option (see Fig.~\ref{figure:new-wire}).
This feature is not communicated through the interface, and requires more steps than would be expected for such a core functionality.
We conclude that Quirk has the least viscous interface for adding primitive gates.

\begin{figure}[t]
    \centering
    \includegraphics[scale=0.4]{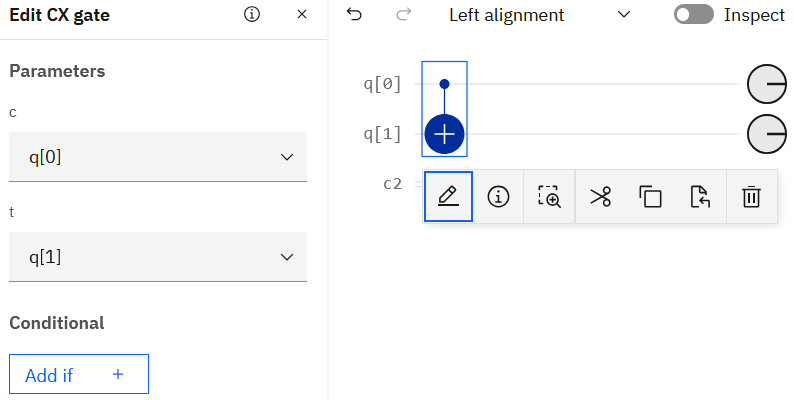}
    \caption{Modifying controls in Qiskit Composer.}
    \label{figure:mod-ctrl}
\end{figure}

\subsubsection{Viscosity and Adding Controls}
In terms of adding a control to a pre-existing gate, both Qiskit Composer and Quirk offer a drag-and-drop interface.
In Quirk, simply dragging a control above or below a gate will automatically connect the control to the gate.
This interface can be used to add classical or quantum controls.
When a control is applicable to a gate, the connection is illustrated graphically to the user by a transparent line and is then applied automatically upon the release of the mouse.
In contrast, Qiskit Composer treats controls as decorators which must be applied to the gate itself.
This is not communicated to the end-user and must be discovered through trial-and-error.
Once a control has been added to the gate, it will appear on the wire immediately above the gate.
The number of their controls, and their positions can be edited using a special window which only appears when double-clicking on a gate (see Fig.~\ref{figure:mod-ctrl}).
We conclude that Quirk has a far less viscous interface for adding controls to gates. 

\subsubsection{Viscosity and Changing Input States}
By default, both Qiskit Composer and Quirk zero-initialize all quantum wires.
This means that all input states are $\ket{0}$ by default.
In OpenQASM 2 there is no way to set the input state of qubit, and consequently Qiskit Composer does not offer an explicit interface to change the input state of a quantum wire.
Instead, the end-user must manually perform state preparation to achieve the desired state.
For example, the $\ket{-}$ state is obtained by applying an $X$-gate followed by an $H$-gate to the chosen wire.
In contrast, Quirk allows an end-user to change the input state of a quantum wire by simply clicking on the input value.
This allows the end-user to choose between an $X$-, $Y$-, or $Z$-basis vector which is useful when testing circuits.
Unfortunately, this mechanism is not clearly communicated to the end-user.
Once the end-user discovers this feature, Quirk has a less viscous interface for changing the input state of a quantum wire.

\subsection{Identification of Situational Factors}
\label{App:Factors}

The following list of questions are provided in~\cite{F2013} to identify situational factors.
We have omitted factor five (the characteristics of the teacher) as the instructors were not yet determined at the time of this analysis.

\vspace{0.5em}
\noindent
\textbf{Specific Context of the Teaching and Learning Situation}
\begin{itemize}
\item ``\emph{How many students are in the class?}''
      Our goal is to admit 10-20 students.
\item ``\emph{Is the course lower division, upper division, or graduate level?}''
      This course could be thought of as a lower divison course, in that it is introductory and provides a survey of quantum computing.
\item ``\emph{How long and frequent are the class meetings?}''
      If we can keep the student's attention on the first day, we will have around 6.5 hours of instructional time.
      This is in addition to 45 minutes of breaks for the students, and an hour long lunch period.
      In contrast to most teaching contexts, we have complete control over how long each of the lessons will be, and the number of lessons.
      This could be broken up into 4-6 lessons, each ranging from 1 hour to 1.5 hours in length.
\item ``\emph{How will the course be delivered: via live classroom instruction, as an online course, or some combination?}''
      One challenge in designing this course is that we would not know whether the course would be hybrid until we finished registration.
      Due to this uncertainty, it is best to work under the assumption that the course will be hybrid, since it is easier to transition from hybrid to live classroom instruction, then it is to add a hybrid component in the last minute.
\end{itemize}

\vspace{0.5em}
\noindent
\textbf{Expectations of External Groups}
\begin{itemize}
\item ``\emph{What does society at large need and expect in terms of the education of these students, in general or with regard to this particular subject?}''
      In general, society would expect our course to supplement these students STEM education.
      In regard to the particular subject, we are in the unique that quantum computing is a new and emerging field for which the general population lacks awareness.
      However, there are some institutions in society which already have expectations.
      For example, the Government of Canada mentions introducing high school students to careers in quantum technology as part of the country's National Quantum Strategy~\cite{ISED2022}.
\item ``\emph{What curricular goals does the institution or department have that affect this course or program?}''
      CISE-Atlantic has asked for us to take a holistic approach to quantum computing, with a focus on how it impacts society.
      The goal is to attract the interest of students who are not intrinsically motivated by STEM topics.
\end{itemize}

\vspace{0.5em}
\noindent
\textbf{Nature of the Subject}
\begin{itemize}
\item ``\emph{Is the subject matter convergent (working towards a single right answer) or divergent (working towards multiple, equally valid interpretations)?}
      Parts of this course are convergent, whereas other parts are divergent.
      For example, given a quantum circuit, there is a single ``correct'' answer as to what it will do to its input qubits.
      On the other hand, there are many circuits which implement the same operator, and several equally valid interpretations of quantum measurement.
      Moreover, questions about the social impact of quantum computing and the best choice of hardware are subjective, and there may be more than one correct answer, provided the answer is adequately defended.
      This could pose an educational challenge, if we do not make it clear to the student that some problems require convergent thinking, while other problems require divergent thinking.
\item ``\emph{Is the subject primarily cognitive or does it involve learning a physical skill as well?}''
      It is cognitive.
\item ``\emph{Is the field of study relatively stable, in a period of rapid change, or in a situation which competing paradigms are challenging each other?}''
      Quantum computing is definitely in a period of rapid change.
      For example, there are many competing paradigms, both in terms of program design and hardware design.
      We should shelter the students from these details, but also make them aware of this ongoing change.
      If we show the students too many software or hardware models, they may get lost in the details and become confused.
      It is interesting to note that there are also competing paradigms in the are of quantum education.
      For example, the conventional approach based in linear algebra and complex analysis has recently been challenged by quantum pictorialism~\cite{DYPPWCKGC2023}.
      This is a graphical approach based on the theory of string diagrams for symmetric monoidal categories.
      This has proven effective for teaching high school students quantum mechanics and quantum computing~\cite{CPYWPCPWSTAKGKC2025}.
      Unfortunately, there is even a lack of consensus among those educators who subscribe to quantum pictoralism, both in terms of which graphical language to use~\cite{LF2023,ERB2020,DYPPWCKGC2023} and whether a matrix-based foundation can be totally supplanted~\cite{SZCFCLHZ2024}.
      Since the studies performed in~\cite{CPYWPCPWSTAKGKC2025} were carried out using the ZX-calculus, this is arguably the safest choice, until further studies are conducted.
\end{itemize}

\vspace{0.5em}
\noindent
\textbf{Characteristics of the Learner}
\begin{itemize}
\item ``\emph{What is the life situation of the students at the moment?}''
      At a very surface level, our audience will be high school students who are living at home with their parents.
      These students will be on summer vacation, but they may have part-time jobs.
      However, we would like to reach out to underrepresented high school students in Nova Scotia.
      This requires a deeper demographic understanding of the life situation of our prototypical student.
      We start by noting that the government of Nova Scotia has designated four employment equity groups: indigenous people; persons with disabilities; African Nova Scotians and other racialized people; and women~\cite{NS2018}.
      Due to their long history of marginalization in Nova Scotia, we have decided to focus our outreach efforts towards the indigenous and African Nova Scotian communities.
      Based on the Unitary Fund's 2025 Quantum Open Source survey, we have also chosen to prioritize Nova Scotian students who identify as women~\cite{UF2025}.
\item ``\emph{What life or professional goals do students have that relate to this learning experience?}''
      Since this is not a mandatory class, then we can safely assume that the content aligns with the personal or career goals of the students (or at least what their parents expect of them).
      However, quantum computing in a diverse field, so it is hard to predict how exactly it relates to their career goals.
      Moreover, we might attract students interested in computer science or physics, meaning that their career goals may only be tangentially related.
      It will be hard to know how the learning experience aligns with the students goals until registration closes.
\item ``\emph{What are their reasons for enrolling?}
      Ideally, students would enroll in this course because they are interested in quantum computing, computer science, physics, or the application of technology to real-world issues.
      Realistically, some students will be enrolled by their parents.
\item ``\emph{What prior experiences, knowledge, skills, and attitudes do the students have regarding this subject?}''
      Based on the Nova Scotia high school curriculum, the students should know basic algebra~\cite{Math9} and have very basic knowledge about science~\cite{Sci8}.
      However, we cannot assume that they have taken courses on matrices, calculus, probability, or physics.
      Moreover, we cannot assume that they have taken a course on computer programming, since this is offered as a grade 12 elective in Nova Scotia~\cite{CS12}.
      We can probably make due without a formal understanding of probability, since~\cite{IBBBDHKKLMPS2023} was able to teach quantum computing using only an intuitive notion of probability.
      While some authors claim that physics is necessary to teach quantum computing~\cite{HITPS2022}, the authors of~\cite{TGRP2022} found that it can be taught to high school students without physics, provided that quantum computing is framed as an abstract model of computation.
\item ``\emph{What are the student’s learning styles?}''
      We do not know since the students have not enrolled yet.
\end{itemize}
These points show that we will not know much about the students prior to the end of registration.
It would be beneficial to collect background information about the students before the event starts, so we can better tailor the material to their skills and interests.

\subsection{Identification of Learning Goals}
\label{App:Goals}

A collection of questions and verbs are provided in~\cite{F2013} to help formulate learning goals with action-oriented language.
Our answers to these questions are listed below.
It should be noted that some answers are hierarchical.
For example, understanding classical logic gates, condition statements, and classical circuits is subsumed by identifying the similarities and differences between quantum computing and classical reversible computing.
These dependencies were taken into account when constructing the final list of learning goals.

\vspace{0.5em}
\noindent
\textbf{Goals Based on Foundational Knowledge}
\begin{itemize}
\item ``\emph{What key information (facts, terms, formula, concepts, relations...) is important for the student to understand and remember in the future (e.g., a year from now)?}''
      \begin{itemize}
      \item Algorithm (term). 
      \item Classical logic gates (concept). 
      \item Conditional statements (concept). 
      \item Classical circuit (concept). 
      \item Qubit (concept). 
      \item Entanglement (concept). 
      \item Quantum gate (term).
      \item Single qubit operations are rotations (concepts).
      \item Quantum computer (term). 
      \item Quantum circuit (concept). 
      \item Quantum teleportation (concept). 
      \end{itemize}
\item ``\emph{What key ideas or perspectives are important for the student to understand in the course?}''
      \begin{itemize}
      \item Computational thinking. 
      \item Uncertainty principle and quantum measurement. 
      \item Quantum computers are real devices which take advantage of unusual physical phenomena. 
      \item Limiting factors for large-scale quantum computing. 
      \end{itemize}
\end{itemize}

\vspace{0.5em}
\noindent
\textbf{Goals Based on Application}
\begin{itemize}
\item ``\emph{What important skills do the students need to learn?}''
      \begin{itemize}
      \item Use Quirk to simulate quantum circuits. 
      \item Read a quantum circuit with basic gates. 
      \item Determine how the inputs to a real-world problem can be mapped to qubits. 
      \item Design a simple quantum circuit based on a detailed problem statement.
      \item Use the ZX-calculus to predict circuit behaviour.
      \end{itemize}
\item ``\emph{What complex projects do the students need to learn how to manage?}''
      \begin{itemize}
      \item Design quantum circuits that work together to solve a single problem.
      \end{itemize}
\end{itemize}

\vspace{0.5em}
\noindent
\textbf{Goals Based on Integration}
\begin{itemize}
\item ``\emph{What connections (similarities and interactions) should the students recognize and make among ideas within this course?}''
      \begin{itemize}
      \item How quantum computing is similar to (reversible) classical computing.
      \item How the principles of quantum mechanics relate to the behaviour of quantum circuits.
      \end{itemize}
\item ``\emph{What connections (similarities and interactions) should the students recognize and make between ideas in this course and in other areas?}''
      \begin{itemize}
      \item How the math students learn in high school can be applied to real problems. 
      \item How technology can interact with the social sciences and humanities, both by solving problems and creating new problems to solve. 
      \end{itemize}
\item ``\emph{What connections (similarities and interactions) should the students recognize and make between ideas in this course and their own lives?}''
      \begin{itemize}
      \item The student should see how applications of quantum computing are directly related to problems in their day-to-day lives. 
      \item The student should be able to relate the basic concepts of quantum mechanics to everyday phenomenon, such as light diffusing through their blinds. 
      \end{itemize}
\end{itemize}

\vspace{0.5em}
\noindent
\textbf{Goals Based on the Human Dimension}
\begin{itemize}
\item ``\emph{What can or should students learn about themselves?}''
      \begin{itemize}
      \item They are capable of doing quantum computing, whether they are interested in the subject or not.
      \item Understand the quantum mechanical phenomenon are complicated, and that it is okay to be confused (even experts get confused).
      \end{itemize}
\item ``\emph{What can or should students learn about interacting with people they may actually encounter in the future?}''
      \begin{itemize}
      \item How to work in small teams to solve problems. 
      \item The importance of collaboration and diversity in both problem solving and scientific research. 
      \item Scientists are just normal people and come from diverse backgrounds.
      \end{itemize}
\end{itemize}

\vspace{0.5em}
\noindent
\textbf{Goals Based on Caring}
\begin{itemize}
\item ``\emph{What changes would you like to see in what students care about, that is, any changes in their values/feelings/interests?}''
      \begin{itemize}
      \item We would like students to learn that STEM can be interesting, even if their primary interests do not lie within STEM. 
      \item We would like students to care about the impact quantum computing can have on the world.
      \item We would like students to appreciate the importance of collaboration and diversity in scientific research. 
      \end{itemize}
\end{itemize}

\vspace{0.5em}
\noindent
\textbf{Goals Based on Learning How to Learn}

\noindent
The questions given in~\cite{F2013} would not make sense for a two-day course.
However, it is still possible to equip students with the skills necessary to help them to learn.
For example, we can introduce students to online programming documentation, so that they are equipped to learn more about quantum programming on their own. 

\subsection{Raw Survey Data}
\label{App:Data}

This section summarizes all of the raw data used in Sec.~\ref{Sect:Eval}.
We begin with the raw demographic data collected from the registration survey.
This question asked students about their race and gender, as is relevant to our outreach efforts.
This information was necessary for CISE-Atlantic to determine the success of our program, so we decide to include it on the registration survey.
We omitted all other demographic questions from the registration survey, since individuals may feel pressured to complete all questions on a registration form, which undermines informed consent.
For this reason, more students responded to this survey than any other question.

\vspace{1em}
\begin{center}
    \scalebox{0.8}{\begin{tabular}{@{}r|lll@{}}
        \toprule
                                & Registered & In-Person & Online \\
        \toprule
        Total Students          & 20         & 8         & 2 \\
        Woman/Nonbinary         & 10         & 6         & 2 \\
        African Nova Scotian    & 4          & 2         & 1 \\
        Indigenous Nova Scotian & 0          & 0         & 0 \\
        Declined to Answer      & 3          & 0         & 0 \\
        \bottomrule
    \end{tabular}}
\end{center}
\vspace{1em}

We collected additional demographic data on the optional pre-event survey provided to all registrants.
Of the $10$ students who responded, $8$ lived in Halifax/Dartmouth, and $9$ had at least one parent who had attended university.
The grade levels of these students at the time of registration (January 2025) are displayed below.
\begin{center}
    \scalebox{0.8}{\begin{tikzpicture}
        \wheelchart[
          data sep=0,
          lines=0.5,
          pie,
          slices style={\WCvarB,draw=black},
          wheel data=\WCperc
        ]{%
          1/yellow/Grade 7,
          1/pink/Grade 8,
          3/lime/Grade 9,
          1/SkyBlue/Grade 10,
          4/Peach/Grade 12
        }
    \end{tikzpicture}}
\end{center}
Of these $10$ students, only $5$ were old enough to have attended \emph{high school level} classes.
Course enrollment data from the remaining $5$ students is displayed below (regular and advanced level courses have been merged).
As mentioned in Sec.~\ref{Sect:Eval}, this table omits IB classes and is therefore incomplete.

\pgfplotstableread[row sep=\\,col sep=&]{
    Course          & Total \\
    Science 10      & 5 \\
    Physics 11      & 2 \\
    Chemistry 11    & 4 \\
    Comp. Prog. 12  & 0 \\
    }\classdata

\begin{center}
    \scalebox{0.8}{\begin{tikzpicture}
        \begin{axis}[
                ybar,
                symbolic x coords={Science 10,Physics 11,Chemistry 11,Comp. Prog. 12},
                extra y ticks={1,2,3,4,5},
                xtick=data,
                xticklabel style={rotate=90},
                nodes near coords,
            ]
            \addplot table[x=Course,y=Total]{\classdata};
        \end{axis}
    \end{tikzpicture}}
\end{center}

Of the $10$ students who completed the pre-event survey, $8$ students also completed the post-event survey.
However, one of these students could not attend the event so their results are omitted.
The following table shows how students responded to certain questions regarding their affinity for STEM and quantum computing on a Likert scale.
It should be noted that the student's responses to the question ``\emph{I enjoyed taking STEM classes in high school.}'' should not change between the pre-event and post-event survey.
Students responses to this question persisted across the pre- and post-event surveys, suggesting that students did not answer these questions randomly.

\vspace{1em}
\begin{center}
    \begin{tabular}{lrlllll}
            \toprule
            Statement & & 1 & 2 & 3 & 4 & 5 \\
            \toprule
            I enjoyed taking STEM   & Before & 0 & 0 & 1 & 2 & 4 \\
            classes in high school. & After  & 0 & 0 & 1 & 2 & 4 \\
            \toprule
            I would like to attend  & Before & 0 & 0 & 0 & 2 & 5 \\
            university.             & After  & 0 & 0 & 0 & 2 & 5 \\
            \toprule
            I would like to work in & Before & 0 & 0 & 3 & 1 & 3 \\
            a STEM-related field.   & After  & 0 & 0 & 2 & 1 & 4 \\
            \toprule
            I would like to work    & Before & 0 & 0 & 5 & 1 & 1 \\
            with quantum computers. & After  & 0 & 1 & 2 & 4 & 0 \\
            \bottomrule
    \end{tabular}
\end{center}
\vspace{1em}

\noindent
We also asked students to identify words related to quantum computing which they recognized, both before and after the event.
The results are displayed below.
Note that the alternative term ``\emph{Mach-Zehnder Interferometer}'' was provided alongside ``\emph{Double Slit Experiment}'' and ``\emph{Born Rule}'' was provided alongside ``\emph{Quantum Measurement}''

\pgfplotstableread[row sep=\\,col sep=&]{
    Term                     & Before & After \\
    Qubit                    & 1      & 7 \\ 
    Bloch Sphere             & 0      & 5 \\
    Superposition            & 1      & 5 \\
    Double Slit Experiment   & 2      & 7 \\
    Entanglement             & 4      & 5 \\
    Quantum Teleportation    & 0      & 4 \\ 
    Quantum Cryptography     & 0      & 1 \\
    Quantum Algorithms       & 0      & 3 \\
    Quantum Machine Learning & 1      & 5 \\
    Quantum Measurement      & 1      & 1 \\
    Schrödinger's Cat        & 3      & 2 \\
    }\termdata

\begin{center}
    \scalebox{0.75}{\begin{tikzpicture}
        \begin{axis}[
                ybar,
                symbolic x coords={Qubit,Bloch Sphere,Superposition,Double Slit Experiment,Entanglement,Quantum Teleportation,Quantum Cryptography,Quantum Algorithms,Quantum Machine Learning,Quantum Measurement,Schrödinger's Cat},
                width=\textwidth/1.5,
                xtick=data,
                xticklabel style={rotate=90},
                nodes near coords,
            ]
            \addplot table[x=Term,y=Before]{\termdata};
            \addplot table[x=Term,y=After]{\termdata};
            \legend{Before,After}
        \end{axis}
    \end{tikzpicture}}
\end{center}

\noindent
General feedback was also collected on the quality of the event, using a Likert scale.

\vspace{1em}
\begin{center}
    \begin{tabular}{llllll}
            \toprule
            Statement & 1 & 2 & 3 & 4 & 5 \\
            \toprule
            I enjoyed the event overall. & 0 & 0 & 0 & 4 & 3 \\
            \toprule
            I enjoyed the single qubit & 0 & 1 & 1 & 1 & 4 \\
            presentation. \\
            \toprule
            I enjoyed the measurement & 0 & 0 & 4 & 1 & 2 \\
            presentation. \\
            \toprule
            I enjoyed the multi-qubit & 0 & 0 & 1 & 4 & 2 \\
            presentation. \\
            \toprule
            I enjoyed the entanglement & 0 & 0 & 1 & 2 & 4 \\
            presentation. \\
            \toprule
            I enjoyed the hands-on activities & 0 & 0 & 1 & 2 & 4 \\
            on day 1. \\
            \bottomrule
    \end{tabular}
\end{center}
\vspace{1em}

\noindent
Of the $10$ students who attended the first day, $5$ students attended the competition.
The following feedback was collected on the quality of the competition, using a Likert scale.

\vspace{1em}
\begin{center}
    \begin{tabular}{llllll}
            \toprule
            Statement & 1 & 2 & 3 & 4 & 5 \\
            \toprule
            I enjoyed the challenges. & 0 & 0 & 1 & 3 & 1 \\
            \toprule
            I learned something from the & 0 & 0 & 1 & 3 & 1 \\
            challenges. \\
            \toprule
            I felt motivated to complete the & 0 & 0 & 3 & 2 & 0 \\
            challenges. \\
            \bottomrule
    \end{tabular}
\end{center}
\vspace{1em}
}{}

\end{document}